# Measuring Systemic Risk: Common Factor Exposures and Tail Dependence Effects

Wan-Chien Chiu[a,*], Juan Ignacio Peña[a], and Chih-Wei Wang[a]

**Abstract**

*We model systemic risk using a common factor that accounts for market-wide shocks and a tail dependence factor that accounts for linkages among extreme stock returns. Specifically, our theoretical model allows for firm-specific impacts of infrequent and extreme events. Using data on the four sectors of the U.S. financial industry from 1996 to 2011, we uncover two key empirical findings. First, disregarding the effect of the tail dependence factor leads to a downward bias in the measurement of systemic risk, especially during weak economic times. Second, when these measures serve as leading indicators of the St. Louis Fed Financial Stress Index, measures that include a tail dependence factor offer better forecasting ability than measures based on a common factor only.*

**Keywords:** *Systemic risk, Tail dependence effects, Correlated jumps, Predictability*
**JEL classification:** *G*01, *G*10, *G*18, *G*20, *G*28

---

We thank David Martínez Miera, Pablo Ruiz Verdú, Pedro Serrano, Sergio Vicente, Silvia Stanescu, and Denisa Banulescu, as well as participants in the 3rd Financial Engineering and Banking Society (FEBS) 2013 Conference on Financial Regulation and Systemic Risk, the 2013 European Financial Management Association (EMFA) annual conference, and the Universidad Carlos III finance seminar, for useful comments. We thank the Editor John A. Doukas for useful suggestions. An anonymous referee provided many astute comments which considerably improved the paper. The usual disclaimer applies. Peña acknowledges financial support from MCI grants ECO2009-12551 and ECO2012-35023.

## 1. Introduction

Multiple published studies document the importance of a stable financial system for not just the financial industry but the real economy as well. Monitoring the whole financial system (not just the banking industry) in turn is required, to guarantee its stability. As the 2007–2012 crises (corporate and sovereign) highlighted, a key factor that affects the stability of the overall financial system, and the real economy, is the level of systemic risk.[1] An accurate measure of this level should be of crucial importance for regulators and investors alike.

In response, extensive literature explores a variety of systemic risk measures (e.g., Bisias *et al.*, 2012). Most measures refer to the aggregate system or individual firm level; in the latter case, systemic risk aggregates can be viewed as the aggregation of financial institutions' risks,[2] which are driven by both common factor exposures to market-wide shocks and additional exposures to other, observed and unobserved factors. A common factor accounts for the systematic component of systemic risk (Das and Uppal, 2004); it cannot capture correlation due to large, infrequent changes (e.g., unexpected failure of a major bank).[3] Therefore, an alternative approach that includes relevant frailty and contagion effects, arising from exposure to unobservable covariates (e.g., common latent factors) is outlined both in Das *et al.* (2007) and Duffie *et al.* (2009). In a frailty setting, the arrival of (bad) news about one firm (extreme negative stock returns) causes a jump in the conditional distribution of hidden covariates and therefore a (negative) jump in any firm's stock returns that depend on the same unobservable covariates.

---

[1] Rajan (2006) highlights the importance of the exposure of the real economy to shocks stemming from the financial sector.

[2] An alternative approximation relies on Lehar's (2005) and Suh's (2012) portfolio approach, which measures systemic risk in the financial sector according to groups of financial firms' risks. Altman and Rijken (2011) similarly assess sovereign default risk by aggregating the Altman's z-scores of non-financial corporations.

[3] Literature on default risks suggests that default times concentrate around periods in which the probability of default of all firms increases. However, this increase cannot be totally, or even partially, explained by firms' common dependence on systematic macroeconomic factors (see Giesecke, 2004;

Unlike previous studies, we use a structural-model approach, rather than a reduced-form approach, and do not make assumptions about the nature of these common factors. Instead, we direct our attention to tail dependence effects that result from simultaneous, extreme equity returns across financial institutions. Furthermore, we focus on the impact of these shocks on systemic risk measures. By adding a correlated jumps factor (as a proxy for tail dependence effects) to the standard Merton (1974) framework,[4] we can address the firm-specific impact of infrequent and extreme events. When a jump occurs, its impact occurs at the same time and in the same direction for all firms (positive or negative), but its size and volatility is specific to each firm. We also refine the methodology proposed by Das and Uppal (2004) to capture

---

Giesecke and Goldberg, 2004; Elsinger *et al.*, 2006a, 2006b).

[4] Adding a correlated jumps factor allows to capture the stylized fact that default correlations may increase when an influential event (e.g. a major bankruptcy), affecting many firms simultaneously, happens. In this vein, Liu, Qi, Shi, and Xie (2013) link default correlation to equity return correlation in the context of the structural framework An alternative view is formulated in Zhou (2001a) which develops a model to compute simultaneous defaults for multiple firms extending the traditional first-passage-time model..

joint tail risk behavior over time. Based on our model, we develop three indicators of systemic stress in the financial industry: (1) *DD*, or the average distance-to-default in a given sector; (2) *NoD*, defined as the number of joint defaults in a given sector; and (3) *PIR*, which is the ratio of the price of insurance against financial distress to the aggregate asset value in a given sector. Given that systemic risk is a multidimensional concept, measures of systemic risk should be based on several relevant characteristics (Bisias et al. ,2012) such as size, interconnectedness, substitutability, leverage, herd effects (clone property), correlation with other sectors of the of the financial industry and correlation with the real economy. Our three measures are attractive because they summarize many of these characteristics, in particular: size, leverage, dependence between firms and the whole stock market, and interconnectedness[5].

In an empirical application, we rely on stock market data, which has a leading role in the price discovery process.[6] Specifically, we focus on the U.S. financial industry and the stock returns of ten largest institutions in four major sectors: depositories, broker-dealers, insurance companies, and others. This concentration on the biggest firms reflects their crucial contribution to systemic risk.[7] The sample period runs from

---

[5] In contrast, measures of systematic risk (e.g. CAPM beta) only take into one of these characteristics, namely the correlation between a firm's stock returns and aggregate market stock returns.

[6] This leading role might entail anticipating trends in subsequent failures (Lehar, 2005) or changes in supervisory ratings four quarters in advance (Krainer and Lopez, 2001). Several articles affirm that equity market information leads the credit risk price discovery process. Zhang *et al.* (2009) observe that credit default swaps are sensitive to jumps in equity returns. document that the equity market leads both the CDS and bond market in the price discovery process see Forte and Peña (2009), and Norden and Weber (2009).

[7] Acharya *et al.* (2010) show that the top six firms in terms of contributions to systemic risk also rank

January 1996 to December 2011. The contribution of this article is threefold. First, our model captures the stylized fact that extreme negative co-movements for large financial institutions are stronger and more frequent in bear than in bull markets. Second, disregarding the impact of tail dependence effects leads to underestimates of the systemic risk level, especially during weak economic times. Third, we analyze whether our systemic risk measures offer leading indicators of alternative measures, using a comparison with a model that includes only common factor effects and a measure based on a public financial stress index, namely, the St. Louis Fed Financial Stress Index (STLFSI).[8] The results show that our measures provide extra forecasting power.

This study extends current literature in several ways. First, to compute systemic risk measures, Lehar (2005) and Suh (2012) consider asset correlations and grant equal weight to both small and large returns. We argue instead that size matters, such that large negative returns must be taken specifically into account to assess the level of systemic risk.[9] Second, traditional jump-diffusion models only allow for individual firm jumps, in terms of both arrival time and size (e.g., Zhou, 2001b), whereas our model assumes a coincident jump arrival time across firms. Third, our study extends among the top seven in terms of total assets. Patro *et al.* (2013) reveals that daily stock return correlations among large financial institutions track with the level of systemic risk. Pais and Stork (2013) suggest that a high stress level in large banks significantly drives systemic instability.

[8] This index is constructed from 18 weekly data series: 7 interest rate series, 6 yield spreads, and 5 other indicators. We chose the STLFSI for three reasons. It is publicly available, spans the whole sample period, and offers the best indicator among U.S. public financial conditional indexes (Aramonte *et al.* 2013).

[9] Bae *et al.* (2003) argue that large negative returns are more influential, and extreme dependence is hidden in traditional correlation measures by the large number of days that present small shocks.

Duffie *et al.*'s (2009) approach; both studies model a firm's default risk, considering observed common factors and unobserved frailty effects, but we employ a different modeling framework and deal with different research goals.[10] Fourth, our study extends the results provided by Acharya *et al.* (2010), who present an expected shortfall model, and the *CoVaR* of Adrian and Brunnermeier (2010). Fifth and finally, we expand on Giesecke and Kim's (2011) model, which is based on a reduced-form framework to consider the influences of market-wide and sector-specific risk factors, as well as spillover effects.

Summing up, our contributions are as follows. First, we propose a new structural-form model that includes exposures to both a common factor exposure and a tail dependence effect. This model effectively captures realistic, time-varying characteristics in extreme stock return correlations, overcoming the limitations of standard models of portfolio credit risk that cannot account for the higher default correlations during tough economic times. Second, the set of alternative systemic risk indicators we propose reflects different perspectives on system-wide stability. Third, our empirical results related to the U.S. market during 1996–2011, we establish three key findings: (1) neglecting tail dependence induces a downside bias in systemic risk measures; (2) considering tail dependence improves a model's forecasting ability; and (3) systemic risk measures based on broker-dealer and insurance sectors lead the public financial stress index on average by a month in advance.

[10] These authors model the frailty effect by including an unobservable macroeconomic variable to determine a firm's default intensity; we consider the frailty effect that results from simultaneous firm-specific shocks in equity markets. Their model is a reduced-form, and ours is a structural-form. Finally, whereas we measure systemic risk in the financial sector, they focus on default clusters among non-financial corporations.

The rest of paper is organized as follows. Section 2, we derive our structural-form model, with both common factor and tail dependence effects. Section 3 contains the methodology and systemic risk measures. After we describe the data in Section 4, we report on the empirical analysis in Section 5. Section 6 concludes.

## 2. The Merton Model with Correlated Jumps

### 2.1. *Asset returns with a common factor and correlated jumps*

This article contributes to emerging literature that proposes bottom-up models of default correlations by modeling the asset value of an individual financial institution, exposed to an observable common factor, tail dependence effects, and an unobservable individual factor. Our model relates to Suh's (2012), which features the common factor with a GARCH process, added to the pure diffusion asset return process. However, we extend this specification by incorporating correlated jumps across individual stocks, which provides a proxy for tail dependence effects. To capture the correlated nature of these jumps, we impose two restrictions. First, we assume that the jump occurs at the same time across all firms. Second, conditional on the jump moving in a given direction (i.e., positive or negative), we assume its size and volatility are firm-specific. With this model, we capture two data features, namely, the correlation between stock returns and a common factor and the infrequent but large changes in stock returns.

Let $V_{j,t}$ and $S_{j,t}$ be firm *j*'s asset value and stock price, respectively, at time *t*. Although $V_{j,t}$ is not observable, it can be inferred from $S_{j,t}$ on the basis of Merton's model. Then let $X_t$ be the common factor. We consider a discrete-time economy for a period of $[0,T]$, where trading takes place at any of $n+1$ trading points $0, \Delta t, 2\Delta t, \ldots, n\Delta t$, and $\Delta t = T/n$. We denote the process of the logarithm of asset return ($v_{j,t} \equiv \log\left(V_{j,t}/V_{j,t-1}\right)$) as follows:

$$v_{j,t} = \mu_j + \delta_j \left(x_t - r\right) + w^*_{j,t}, \quad (1)$$

$$w^*_{j,t} = w_{j,t} + Q_j N\left(\Delta t\right) - \bar{Q}_j \lambda, \quad (2)$$

$$w_{j,t} \sim N\left(0, \xi_j\right), \quad (3)$$

where $\mu_j$ represents the long-run mean of firm *j*'s log-return, $x_t$ is the log-return of the common factor, *r* is the risk-free interest rate, and $w^*_{j,t}$ indicates exposures to other factors. To capture the impact of correlated jumps across firms' assets, we partition $w^*_{j,t}$ into two components in Equation (2): $w_{j,t}$ is an idiosyncratic factor that follows a multivariate distribution without considering extreme dependence,[11] and $Q_j N\left(\Delta t\right)$ and the adjustment term $-\bar{Q}_j \lambda$, which account for the tail dependence exposure term.[12]

---

[11] The specification of $w_{j,t}$ will be described in the following section.

[12] We subtract $\bar{Q}_j \lambda$, where $\bar{Q}_j = E\left[Q_j\right]$, to impose a zero mean Poisson process.

With this term, the firm's asset value can jump when its equity price suddenly suffers a large movement, due to the arrival of news. For example, extreme stock returns for one firm may cause a jump in the conditional distribution of hidden covariates, leading to a jump in the stock returns of other firms whose stock returns depend on the same unobservable covariates

In our effort to model large changes in prices occurring at the same time across firms' asset returns, we assume that the arrival of jumps is coincident across all firms' asset returns; that is, $N_j(\Delta t) = N(\Delta t)$, such that $N(\Delta t)$ is the standard Poisson counting process with mean and variance $E(N(\Delta t)) = \lambda = Var(N(\Delta t))$. We denote $Q_j$ as a random jump amplitude on the log-return if the Poisson event occurs. Furthermore, we let $Q_j$ and $N(\Delta t)$ be mutually independent; $Q_j N(\Delta t)$ is a Poisson random sum of normal random variables. Therefore,

$$Q_j N(\Delta t) = \sum_{k=1}^{N(\Delta t)} Q_j^{(k)}(\Delta t), \quad (4)$$

where $Q_j^{(k)}(\Delta t) \sim N(a_j, b_j^2)$ for $k = 1, 2, \ldots$. In this setting, the distribution of the jump size is asset-specific in its mean and volatility, but the jump arrives at the same time for all firms. For our model, a realization of one Poisson process triggers simultaneous, large movements across multiple companies.

Noting the dynamics of the common factor, we employ a GARCH-type model. Specifically, we follow Heston and Nandi (2000) and model the common factor, under the physical measure P, as

$$x_t = r + \lambda^P h_t + \sqrt{h_t}\varepsilon_t, \ (5)$$

$$h_t = \omega + \alpha\left(\varepsilon_{t-1} - \gamma\sqrt{h_{t-1}}\right)^2 + \eta h_{t-1}, \ (6)$$

where $r$ is the continuously compounded interest rate for the interval between $t$ and $t - \Delta$, $\varepsilon_t$ is a standard normal disturbance, and $h_t$ is the conditional variance of the log-return between $t$ and $t - \Delta$.[13] The conditional variance of an asset return is time varying, i.e.,

$$Var\left(v_{j,t} \mid \varphi_{t-1}\right) \equiv \sigma^2_{j,t} = \delta^2_j h_t + \xi_j + \lambda \hat{b}^2_j, \ (7)$$

where $\hat{b}^2_j = a^2_j + b^2_j$. We provide the derivation in Appendix A.

2.2. *Structural-form model with a factor jump diffusion process*

We define equity $S$ under the risk-neutral measure (RN) as a call option with maturity $T$:

$$S_{j,t} = e^{-r(T-t)} E^{RN}\left[\max\left(V_{j,T} - D_{j,T}, 0\right)\right], \ (8)$$

such that $S_{j,t}$ denotes the equity price of firm $j$ at time $t$. Following Duan (1995) we assume that the RN measure satisfies the locally risk-neutral valuation relationship (LRNVR), in which the expected return under the RN measure is the risk-free rate, but the one-period-ahead conditional variance of the return stays the same under the $P$ and RN measures. Adopting the same assumption, Heston and Nandi (2000) show that under the RN measure,

$$x_t = r - \frac{1}{2}h_t + \sqrt{h_t}\varepsilon_t, \quad (9)$$

$$h_t = \omega + \alpha\left(\varepsilon_{t-1} - \left(\gamma + \lambda^P + \frac{1}{2}\right)\sqrt{h_{t-1}}\right)^2 + \eta h_{t-1}. \quad (10)$$

[13] We make $r$ constant for a certain time, using the mean of the risk-free interest rate, that is, the 1-year Treasury constant maturity rate obtained from the U.S. Federal Reserve, divided by 252.

Heston and Nandi (2000) also derive the following conditional generating function of the future common factor:

$$f(\phi) \equiv E_t\left[X_T^\phi\right] = X_t^\phi \exp\left(A(t;T,\phi) + B(t;T,\phi)h_{t+1}\right), \quad (11)$$

where the coefficients are recursively determined as:

$$A(T;T,\phi) = 0, \quad (12)$$

$$A(t;T,\phi) = A(t+1;T,\phi) + \phi r + B(t+1;T,\phi)\omega - \frac{1}{2}\ln\left(1 - 2\alpha B(t+1;T,\phi)\right), \quad (13)$$

$$B(T;T,\phi) = 0, \quad (14)$$

$$B(t;T,\phi) = \phi\left(\lambda^P + \gamma\right) - \frac{1}{2}\gamma^2 + \eta B(t+1;T,\phi) + \frac{\frac{1}{2}(\phi-\gamma)^2}{1\text{-}2\alpha B(t+1;T,\phi)}. \quad (15)$$

Accordingly, we can derive the conditional generating function for asset values. We start by noting that under the RN measure,

$$\log\frac{V_{j,T}}{V_{j,t}} = \left(r - r\delta_j - \bar{Q}_j\lambda\right)(T-t) + \delta_j \log\frac{X_T}{X_t} + W_{j,t}^T + Q_j N(T), \quad (16)$$

where $W_{j,t}^T \equiv w_{j,t+\Delta t} + \ldots + w_{j,t+n\Delta t}$ and $N(T) = N(\Delta t) + N(2\Delta t) + \ldots + N(n\Delta t)$.[14]

Thus we know

---

[14] Bates (1991) shows that the difference between the risk-neutral and true parameters of $Q_j$ and $N$ is small, both qualitatively and quantitatively. Thus, we assume $\bar{Q}_j$ obtained under physical probability is the same as that obtained under risk-neutral probability.

$$V_{j,T}^{\phi}=V_{j,t}^{\phi}X_t^{-\delta_j\phi}e^{\phi\left(r-r\delta_j-\bar{Q}_j\lambda\right)(T-t)+\phi W_{j,t}^{T}+\phi Q_j N(T)}X_T^{\delta_j\phi}, \quad (17)$$

and we can derive the conditional generating function for asset values:

$$g_j(\phi)\equiv E_t\left[V_{j,T}^{\phi}\right]=V_{j,t}^{\phi}X_t^{-\delta_j\phi}e^{\phi\left(r-r\delta_j-\bar{Q}_j\lambda\right)(T-t)+\phi^2\xi_j(T-t)/2}f\left(\delta_j\phi\right)E_t\left[e^{\phi Q_j N(T)}\right], \quad (18)$$

where $E_t\left[e^{\phi Q_j N(T)}\right]=\exp\left(\lambda(T-t)\left(\exp\left(a_j\phi+\frac{1}{2}b_j^2\phi^2\right)-1\right)\right)$. Appendix B contains further details. With the assumption that equity is valued as a European call option, we determine the equity valuation formula:

$$\begin{aligned}S_{j,t}&\equiv e^{-r(T-t)}E_t^{RN}\left[\max\left(V_{j,T}-D_{j,T},0\right)\right]\\&=\frac{1}{2}V_{j,t}+\frac{e^{-r(T-t)}}{\pi}\int_0^{\infty}\mathrm{Re}\left[\frac{D_{j,T}^{-i\phi}g_j^*(i\phi+1)}{i\phi}\right]d\phi-D_{j,t}\left(\frac{1}{2}+\frac{1}{\pi}\int_0^{\infty}\mathrm{Re}\left[\frac{D_{j,T}^{-i\phi}g_j^*(i\phi)}{i\phi}\right]d\phi\right),\end{aligned} \quad (19)$$

where $g_j^*(\cdot)$ comes from $g_j(\cdot)$, by replacing $\lambda^P$ with –1/2 and $\gamma$ with $\gamma^*\left(\equiv\gamma+\lambda^P+1/2\right)$.[15]

## 2.3. *Dynamics of individual factors*

The unobservable individual factors $w_{j,t}$ may be correlated across firms and over time. In particular, we assume that the vector of individual factors $\mathbf{w}_t\equiv\left[w_{1,t}\cdots w_{N,t}\right]'$ follows a multivariate normal distribution with a time-varying covariance matrix,

$$\mathbf{w}_t\sim MVN\left[\mathbf{0},\mathbf{\Omega}_t\right], \quad (20)$$

where the (*j*,*k*) element of $\mathbf{\Omega}_t$ is $\xi_{jk,t}$. Then we apply the dynamic conditional correlation (DCC) model (Engle, 2002) to estimate the time-varying asset return

[15] The debt is assumed to grow at the risk-free interest rate (Lehar, 2005).

correlations of idiosyncratic components for the dynamics of $\mathbf{\Omega}_t$.[16]

To estimate the time-varying covariance matrix $\mathbf{\Omega}_t$, we first use the estimates of $\hat{\Theta}_j$ for institution $j$ to estimate the time series $\{V_{j,t}\}$ and $\{v_{j,t}\}$ then obtain the residuals $\hat{w}_{j,t}$, defined as:

$$\hat{w}_{j,t} \equiv v_{j,t} - \left(\hat{\mu}_j + \hat{\delta}_j \left(x_t - r\right) + \left(Q_j N\left(\Delta t\right) - \bar{Q}_j \lambda\right)\right). \quad (21)$$

2.4. *Estimation*

The parameter estimation proceeds in three steps. First, we estimate the common factor parameters $\{\omega, \alpha, \eta, \gamma\}$ in the system of Equations (5) and (6), using the maximum likelihood method, according to the common factor data series. Second, we identify $\lambda$, $a_j$, and $b_j$, similar to the way Das and Uppal (2004) do.[17] Third, we make two assumptions regarding the estimation of the parameters related to the asset return process of individual institutions. That is, we assume that the maturity of the implied call option is one year, in line with previous literature (e.g., Ronn and Verma, 1986; Lehar, 2005; Suh, 2012). Using the sum of half of the long-term debt plus the short-term debt, we proxy for the debt amount $D_{j,t}$ within the assumed maturity of one year, in accordance with KMV's methodology. For consistency with prior literature (e.g.,

[16] In contrast, Suh (2012) features the correlation of individual factors based on diagonal VECH, and Lehar (2005) uses an exponentially weighted moving average scheme. We prefer DCC over other types of multivariate volatility process models (e.g., Brownlees and Engle, 2011) because they are easier to estimate and their parameters have an intuitive interpretation , see Silvennoinen, and Teräsvirta (2009).

[17] The correlated jump intensity derives from stock market information. As in Das and Uppal (2004), we assume a jump diffusion process for the stock return process, and we estimate the parameters by minimizing the root mean square error (RMSE) of two metrics, based on co-skewness and excess kurtosis.

Duan, 1994, 2000),[18] we used historical returns to estimate the parameters. For one institution at a time, with maximum likelihood methods, we estimated the parameters $\Theta_j = \{\mu_j, \delta_j, \xi_j\}$ for an institution *j*'s asset return. Given institution *j*'s equity price and debt data $\mathbf{S}_j = \left[S_{j,1} \ldots S_{j,n}\right]'$, $\mathbf{D}_j = \left[D_{j,1} \ldots D_{j,n}\right]'$, and common factor data $\mathbf{x} = \left[x_1 \ldots x_n\right]'$, we derive the following log-likelihood function:

$$\begin{aligned}\log L\left(\Theta_j \mid \mathbf{S}_j, \mathbf{x}, \mathbf{D}_j\right) = &-\frac{n-1}{2}\log\left(2\pi\right) - \sum_{t=2}^{n}\log V_{j,t} - \frac{1}{2}\sum_{t=2}^{n}\log\sigma_{j,t}^2 - \sum_{t=2}^{n}\log\left(\frac{\partial S_{j,t}}{\partial V_{j,t}}\right) \\ &-\frac{1}{2}\sum_{t=2}^{n}\frac{\left\{v_{j,t} - \left(\mu_j + \delta_j\left(x_t - r\right) + a_j\lambda - \bar{Q}_j\lambda\right)\right\}^2}{\sigma_{j,t}^2}\end{aligned}, \quad (22)$$

$$\begin{aligned}\frac{\partial S_{j,t}}{\partial V_{j,t}} = &\frac{1}{2} + \frac{e^{-r(T-t)}}{\pi}\frac{1}{V_{j,t}}\int_0^\infty \mathrm{Re}\left[\frac{\left(D_{j,t}e^{r(T-t)}\right)^{-i\phi}\left(i\phi+1\right)g_j^*\left(i\phi+1\right)}{i\phi}\right]d\phi - \frac{D_{j,t}}{\pi V_{j,t}} \\ &\times\int_0^\infty \mathrm{Re}\left[\left(D_{j,t}e^{r(T-t)}\right)^{-i\phi} g_j^*\left(i\phi\right)\right]d\phi.\end{aligned} \quad (23)$$

Here, $V_{j,t}$ and $\sigma_{j,t}$ provide the solutions to Equations (19) and (7), and $v_{j,t}$ represents the log return of $V_{j,t}$.

## 3. Methodology and Systemic Risk Measures

In the following section, we use a model that only accounts for exposure to the common factor as a benchmark. Our model thus nests the benchmark model when $\lambda = 0$. We compute risk indicators from both our proposed and the benchmark model, using the following methodological procedure.

### 3.1. *Monte Carlo simulation*

[18] We use the 1-year Treasury constant maturity rate obtained from the U.S. Federal Reserve as the risk-free interest rate.

We employ Monte Carlo Simulation because no analytical solution is available for the systemic risk measures over a multi-period time horizon. We draw standard normal random variables and simulate a hypothetical future common factor realization, according to Equations (5) and (6). Next we generate the random variable of correlated jumps by drawing from normal random variables, with a pre-specified mean and standard deviation of firms' jump magnitudes, as well as a Poisson random variable with the pre-specified intensity $\lambda$. Finally, we draw multivariate normal random variables as specified by Equation (20) and repeat the process 10,000 times.

3.2. *Rolling Windows*

With a rolling window approach, we consider the extent to which systemic risk measures vary over time, such that we can avoid look-ahead bias. Our one-year rolling window updates every month. Thus, we construct a subsample for month $t$, using the information from months $t$, $t-1$, $t-2$, …, $t-11$. We repeat this calculation for month $t+1$, rolling the sample one month forward. For example, the first subsample, corresponding to December 1996, contains data from January 1996 to December 1996. The sample gets updated by including the following month and discarding the first one, so the second subsample would correspond to January 1997 and contain data from February 1996 to January 1997. Monthly updating effectively balances accuracy against the computational burden.

### 3.3. *Systemic risk measures*

Extant literature offers a plethora of measures of systemic risks (for a review, see Rodriguez-Moreno and Peña, 2013). Such measures should detect at least two kinds of situations and cover two different dimensions. First, some measures warn of the persistent build-up of imbalances within the financial sector (using monthly or quarterly data), whereas others capture the abrupt materialization of systemic risk (daily or intraday data). Second, there should be measures based on the aggregate market level (e.g., interbank rates, stock market, CDS indexes), as well as measures at the individual institution level. No single measure is "best," and alternative measures may be devised according to the objectives of the systemic risk analysis. Our model specifies the dynamics pertaining to both individual institutions and their tail-risk connection, so it supports the calculation of a wide range of systemic risk measures. We develop three alternative indicators.

(1) *DD*: the average distance-to-default in a given sector over a fixed time horizon

*T*he *DD* has been used as proxy for identifying a financial sector's stability. For example, Jokipii and Monnin (2013) and Carlson *et al.* (2011) both use *DD* to signal distress in the financial sector; the former finds a positive link between this measure and real output growth, especially during periods of instability, and the latter suggests that *DD* offers a leading indicator of real economic activity (e.g., bank lending standards and terms).[19] For this study, we compute *DD* using a structural form model, with and without jump effects. In line with the Merton's *DD* framework, it entails the logarithm of asset value minus the logarithm of debt value, divided by the standard deviation of this difference. Formally,

$$DD \equiv \frac{E\left[\ln V_T - \ln D_T\right]}{Std\left[\ln V_T - \ln D_T\right]}, \quad (24)$$

where $V_T$ and $D_T$ are the asset's market value and the debt's face value, with maturity T.[20] At a given time point *t*, for every firm *j* in a given sector, we compute the daily simulated asset values for the next six months, generated by Monte Carlo simulation. Then we average the difference between the log-asset value and log-debt value and use the result as the numerator; the standard deviation of this difference serves as the denominator. Finally, we compute the average sector value as the weighted-average of all firms in a given sector, with weights based on asset size.[21] The lower the *DD* measure, the higher the level of systemic risk.

---

[19] We should point out that our paper's results are based on portfolios of financial institutions. Random measurement errors in the degree of indebtedness of individual institutions tend to be compensated for with a portfolio approach (Saldias, 2013). Moreover, aggregate *DD* is a valuable tool for monitoring risk profiles in the financial sector, despite the modeling assumptions inherent to a Merton-based model (Gropp *et al.*, 2009; Vassalou and Xing, 2004).

[20] The formula of our *DD* measure is consistent with the general form in the Merton model. In a standard Merton *DD*, $lnV_T$ has a mean of $E\left[\ln V_T\right] = \ln V_0 + \left(\mu_V - 0.5\sigma_V^2\right)T$, a standard deviation of $Std\left[\ln V_T\right] = \sigma_V\sqrt{T}$, and a normal distribution. The *DD* in Merton's model is $\left[\left(\ln\left(V_0/D_T\right) + \left(\mu_V - 0.5\sigma_V^2\right)T\right)/\sigma_v\sqrt{T}\right]$. Because $D_T$ is constant, the numerator can be represented as $\ln\left(V_0/D_T\right) + \left(\mu - 0.5_V^2\right)T = E\left[\ln V_T - \ln D_T\right]$, and the denominator can be rewritten as $\sigma_V\sqrt{T} = Std\left(\ln V_T\right) = Std\left[\ln V_T - \ln D_T\right]$.

[21] We assume that the largest institutions should contribute strongly to overall systemic risk in the financial system.

(2) *NoD*: the number of joint defaults in a given sector over a fixed time horizon.

If a significant number of financial firms default at the same time, the whole financial system (through asset-fire sale or network contagion) might be severely affected (Lehar, 2005). A financial institution is in default if the market value of its assets falls below of the face value of its debt within the next six months. Thus at a given time point *t* and for every firm *j* in a given sector, we compute daily simulated asset values for the next six months, generated by Monte Carlo simulation. Then we compare firm *j's* asset value against the face value of its debt. If the latter is higher than the former, firm *j* is in default; we compute the number of defaulted firms for each sector. The larger *NoD*, the higher the level of systemic risk.

(3) *PIR*: the ratio of the price of insurance against financial distress to the aggregate asset value in a given sector.

This systemic risk measure, proposed by Huang *et al.* (2009), is associated with the idea of assessing the systemic risk of the financial sector by computing the price of the government's contingent insurance against large default losses in the financial sector. With our structural-form model, we consider the amount of financial institution debt that cannot be covered by the institutions themselves, as proxy for this insurance, which we refer to the price of insurance (*PI*). The economic intuition backing this measure is that it proxies the theoretical premium of a risk-based deposit insurance scheme guaranteed by the government (as an insurer of last resort) covering losses exceeding banking sector's total assets.

We measure *PI* by computing a put option value based on the Merton's framework, as Lehar (2005) does. Formally, the price of insurance $PI_t^j$ of a firm *j* at time *t* for a horizon of *T* is $e^{-r(T-t)} \times E\left[\max\left(D_T^j - V_T^j, 0\right)\right]$, where $D_T^j$ is the face value of the firm's debt at time *T*, and $V_T^j$ is the market value of the firm's assets at time *T*. We also consider sector-wide distress, equal to the ratio of the sector's *PI* values to the sector's total asset value over the next six months. We call this risk measure *PIR* and compute it using the formula $PIR_t = \sum_j PI_t^j \Big/ \sum_j Asset_t^j$. Intuitively, the higher the *PIR*, the higher the systemic risk level.

In summary, the indicators rely on intuitive economic interpretations, and we use them to illustrate the temporal trend of overall systemic risk levels. In particular, *DD*, *NoD*, and *PIR* are attractive because they summarize key determinants of systemic risk (firms' size, firms' leverage, dependence between firms and the whole market) as suggested by Acharya *et al.* (2010); they also reflect interconnectedness, as suggested by Cummins and Weiss (2010) and Jobst (2012).[22] We repeat the Monte Carlo simulation procedure for each month from December 1996 to December 2011, yielding

[22] Cummins and Weiss (2010) suggest three primary indicators of systemic risk: (1) size, (2) interconnectedness, and (3) lack of substitutability. Also, Jobst (2012) relates short-term liquidity risk to size and interconnectedness.

monthly time series for each measure.

## 4. Data

### 4.1. *Sample selection*

Our sample comprises large, U.S. financial institutions and spans January 1996 to December 2011. We choose firms with available daily equity prices and quarterly balance sheet information in the CRSP and COMPUSTAT databases.[23] We lag all accounting information by three months to acknowledge reporting delays and substitute for any missing accounting data with the most recent prior observation. The quarterly accounting data is linearly interpolated between quarterly reporting dates at daily frequency. Firms constitute four groups (Acharya *et al.*, 2010; Brownlees and Engle, 2011): depositories, brokers-dealers, insurance companies, and others.[24] We use daily equity returns given that jumps probably appear more clearly in high frequency data.[25] We select the biggest firms based on their book value of total assets at the starting date of each estimation sample for each sector at a given time. Furthermore, the sample only contains firms continuously listed in a prior year, to ensure perfect matches in the number of observations at firm-level and system-level. To avoid survivorship bias, merged or bankrupt entities are also included in the sample, as long as their equity and balance sheet information are available. For each month and in each given sector, the sample includes the ten largest firms. Specific names may change over time because of bankruptcies, mergers, or other reasons. Our sample contains 25 depositories, 24 broker-dealers, 22 insurance companies, and 31 other firms. For depositories, broker-dealers, and others, the average number of changes in the identities of the top ten each month is roughly 0.2, or 2.5 per year. For insurance companies, the average is 1 firm per year. In addition to usual mergers and acquisitions,[26] the reasons for these changes relate to financial distress or bankruptcy (e.g., filing for Chapters 7 or 11). The numbers of bankrupt firms across sectors are as follows: 1 of 25 depositories; 1 of 24 broker-dealers; 0 of 22 insurance companies; and 4 of 31 others.[27]

[23] We collect information about daily equity prices and returns, as well as outstanding shares, from CRSP. We obtain information about total assets, debt in current liability, long-term debt due in one year, and outstanding shares (if missing in CRSP) from COMPUSTAT.

[24] The four groups are depositories (two-digit standard industrial classification [SIC] code 60); brokers-dealers (four-digit SIC code 6211); insurance companies (two-digit SIC code 63 or 64), and others (two-digit SIC codes 61, 62 except 6211, 65, or 67). We assigned Goldman Sachs to the broker-dealers group, despite its SIC code of 6282, following Acharya *et al.* (2010).

[25] Lehar (2005) and Suh (2012) use lower frequency data (monthly and weekly).

### 4.2. *Monthly-interval observations*

By moving the estimation window month by month, we obtain time-varying estimated parameters and risk measures at the end of each month, from December 1996 to December 2011. This sample contains 181 monthly observations for each parameter and measure. Appendix C provides descriptions of the firms in the empirical application. We compute SIZE and LVG (leverage), both at firm and sector-level, at time $t$. The former is the logarithm of the book value of total assets (firm-level) and the logarithm of the summation of all firms in a sector (sector-level); the latter is the quasi-market value of assets, divided by the market value of equity (firm-level) and the weighted average leverage (sector-level), with weights based on market equity.[28]

[26] For example, Bear Stearns was acquired by JPMorgan Chase in 2008.
[27] Bankrupt firms are Washington Mutual Inc., Lehman Brothers Holdings Inc., Finova Group Inc., MF Global Holdings Ltd., New Century Financial Corp., and Thornburg Mortgage Inc.

Figure 1 shows the annual returns across sectors and for the CRSP value-weighted index, which we use to capture the common factor. The sector-level annual returns, ending at month *t* for sector *k*, can be calculated by $r_{k,t} = \sum_{j=1}^{10} w_{j,k,t} \times r_{j,k,t}$, where $r_{j,k,t}$ is firm *j*'s annual return, and $w_{j,k,t}$ is the weight based on market equity for firm *j* at the end of month *t*.

We observe a similar pattern across industries. All sectors show positive performance from 1996 until the end of 1998, when the LTCM crisis occurred. Recovery was slow until the bursting of the dot.com bubble in March 2000. Then a subperiod, until 2003, featured momentum toward recovery. Between mid-2005 and mid-2007, all sectors indicated positive performance, until distress symptoms appeared around July 2007, at the start of the subprime crisis. The market bottomed around March 2009, with a strong rebound in mid-2009. The market plunge around May 2010 led to no clear recovery signals until the end of 2011. Notice that the others sector's stock returns seemed more volatile than the three named sectors, though the returns of the various sectors mimic the overall market trend, just with more volatility.

[28] Following Acharya *et al.* (2010), LVG is the standard approximation of leverage, where quasi-market value of assets is obtained from the book value of assets, minus the book value of equity and plus the market value of equity.

**[Insert Figure 1 Here]**

Table 1 contains the summary statistics by sector. In terms of size, we find no clear differences across sectors. Leverage is highest for the broker-dealers sector (12.26), followed by insurance (11.93) and others (11.22); the least leveraged sector by far was depositories (7.94). The best return/risk ratio accrues to the brokers-dealers (0.54), followed by insurance companies (0.35), depositories (0.34), and then others (0.32). We classify risk measures using the subindex "ben" to refer to benchmark-based measures (i.e., accounting for common factors only). Measures without this subindex reflect the full model (common factor plus tail dependence effects). Because the *DD* ($DD_{ben}$) indicates the distance to default over the next six months, lower value implies higher systemic risk for a sector. In the broker-dealers sector, this measure comes closest to default, with an average value of 2.59 (6.01), followed by others at 3.45 (5.51), depositories with 6.31 (8.0), and finally insurance companies at 10.99 (12.18). Furthermore, *NoD* ($NoD_{ben}$), or the number of defaults among the 10 biggest financial institutions, achieves the highest values in the others sector at 2.38 (1.42), followed by broker-dealers with 2.26 (0.96). Both depositories with 0.79 (0.26) and insurance companies with 0.36 (0.23) exhibit fewer defaults. Finally, for *PIR* ($PIR_{ben}$), the ratio of a sector's price of insurance against financial distress to the sector's total assets, others sector reveals the largest value of 39.90 (15.07), followed by broker-dealers with 22.22 (2.61); depositories and insurance companies again indicated lower values, of 4.50 (0.34) and 3.75 (1.42), respectively. These measures accordingly indicate that the riskiest sectors are broker-dealers and others, followed by depositories and insurance companies. In all cases, the measure from the full model indicates more systemic risk than a measure based on the benchmark.

**[Insert Table 1 Here]**

Regarding the correlations across measures, *DD* reveals negative correlation with *NoD* (–0.69) and *PIR* (–0.42), whereas *NoD* and *PIR* indicate a positive correlation (0.75). We estimate correlations across the four sectors for each measure too and find that they vary. For example, the highest correlation arises between depositories and broker-dealers for *DD* and *PIR*, as well as between depositories and insurance companies for *NoD*. Correlations across sectors for *PIR* generally are greater than those for the other two measures. The correlations range from 0.44 to 0.82 for *DD*, from 0.41 to 0.87 for *NoD*, and from 0.74 to 0.88 for *PIR*.[29]

## 5. Empirical Analysis

With our empirical analysis, we explore the effect of combining two factors (common factor and tail dependence effects) to measure systemic risk. Therefore, we first document estimation results from the correlated jumps and structural-form models. In the next section, we present a preliminary comparison between the full model and benchmark-based measures, then test whether our full model systemic risk measures constitute leading indicators of benchmark-based ones and of the St. Louis Fed Financial Stress Index (STLFSI).

[29] Detailed information about the correlations for each measure within the four sectors is not reported here but is available on request.

5.1.*Estimation results*

5.1.1. *Tail dependence parameters*

To characterize the sector-level behavior of the tail dependence effects, which we proxy for with correlated jumps, we average the firm-specific estimates into one single measure for the mean and the volatility of the size of the correlated jumps by sector, denoted *mu_coj* and *std_coj*. With a rolling window approach, we compute the time series for $\lambda$, *mu_coj*, and *std_coj*.[30] These estimates describe the properties of simultaneous shocks in the equity market. Figure 2 reports three time-varying variables from 1996 to 2011, by sector.[31]

For $\lambda$, depositories and insurance companies indicated similar, smooth moving behaviors: usually below 0.1 before 2006, increasing during 2007, peaking (around 0.3) and staying high for a while, dropping to a pre-crisis level in mid-2009, and increasing again in mid-2011. For broker-dealers, the parameter moved steadily, with low levels before 2005, slight increases in the following two years, and a peak (0.2) near the time of Lehman's failure. After that event, it dropped to a pre-Lehman level, though it appeared more unstable than it is before 2005. Finally, the $\lambda$ of the others sector fluctuated more frequently before 2006, then reached its peak in the fourth quarter of 2008 (>0.35) and remained at a relatively high level (>0.15) for a longer time during 2008–2009 than the three named sectors. Again we observe a clear increase after mid-2011. Overall then, the intensity of correlated jumps across sectors began to increase before the subprime loan crisis of 2007, reached its peak around the time Lehman failed, decreased, and then increased again in mid-2011, coincident with the Eurozone crisis. This evidence suggests that the probability of simultaneous jumps is higher during crises.

[30] For example, in the case of the parameter $\lambda$, we estimate it for a group of the ten largest financial institutions in each sector, using data from January 1996 to December 1996, and assign the calculated value to December 1996. Then we repeat the procedure using data from February 1996 to January 1997 and assign this calculated value of $\lambda$ to January 1997, and so on.
[31] Specific information about the main systemic events from 2007 to 2011 is available at http://timeline.stlouisfed.org/.

The average jump size *mu_coj* is close to zero throughout the sample period for insurance companies; it was not in the other three sectors. We identified negative jumps around the time of Lehman's bankruptcy for broker-dealers, others, and depositories, with average sizes of –0.10, –0.07 and –0.05, respectively. The others sector also suffered negative jumps in the first half of 2009, possibly due to events related to the crisis and subsequent bailout of Fannie Mae and Freddie Mac; this sector contains many firms involved in mortgage markets.[32]

For the jump volatility *std_coj*, the behavior appeared similar across sectors: constantly below 0.05 and very stable until the end of 2007, increasing at the beginning of 2008, reaching historically high levels around mid-2009, and dropping to lower levels thereafter.

This collected evidence matches our intuition regarding the model parameters. In most cases, $\lambda$ and *std_coj* are higher and *mu_coj* displays negative values during episodes of systemic risk. That is, acute stress situations in the financial industry are coincident with higher frequencies of simultaneous, negative, extreme jumps in the stock returns of firms in that industry.

**[Insert Figure 2 Here]**

To examine whether the correlated jumps display specific behavior during the 2007–2009 crisis, we analyze results for 2005–2011 and compare the estimates across three periods: pre-crisis (July 2005–June 2007), crisis (July 2007–June 2009), and post-crisis (July 2009–June 2011). The results in Table 2 are specific to depositories, broker-dealers, insurance companies, and others (Panels A–D, respectively) and indicate significant differences between the pre-crisis and crisis periods for each group. As expected, the intensity and volatility of correlated jumps are higher in the crisis period; and the mean of correlated jumps is strongly negative in this period.

[32] On March 11, 2009, Freddie Mac announced net losses of $23.9 billion for the fourth quarter of 2008 and $50.1 billion for 2008 as a whole. Its conservator submitted a request to the U.S. Treasury Department for an additional $30.8 billion in funding, under a Senior Preferred Stock Purchase Agreement.

**[Insert Table 2 Here]**

The others sector always reveals the highest $\lambda$. In the crisis period, it also has the highest value for *std_coj* (0.13) and the lowest *mu_coj* (–0.02), followed by broker-dealers (–0.01). Thus, in this others sector, negative shocks are deeper and more frequent, and their size is more volatile. We also note significant increases in $\lambda$ during the crisis period. For example, for depositories it increased more than threefold compared with the pre-crisis period (0.13 versus 0.04), doubled in size (0.06 versus 0.03) for broker-dealers, and increased notably (0.19 versus 0.10) in the others sector. The statistical tests thus support our intuition that there was a higher probability of simultaneous negative shocks in the equity market during the 2007–2009 financial crisis, compared with both preceding and posterior periods.

5.1.2. *Common factor parameters*

The parameters of the common factor component (benchmark model) are $\mu$, $\delta$, and $\xi$, which capture the long-run mean of asset returns, exposure to the common factor, and variance in idiosyncratic factors, respectively. We average firm-level estimates to obtain sector-level variables, and to distinguish the estimates of the full model from those of the benchmark model, we use notations of $\mu__{ben}$, $\delta__{ben}$, and $\xi__{ben}$, for the latter. Table 3 contains the estimates by sector, along with the results of the mean tests for estimates derived from the full model and from the benchmark. First, we observe that both $\delta$ and $\xi$ are significantly lower than $\delta_{_ben,}$ and $\xi_{_ben}$ (see Columns 3, 6, and 9 in Table 3). By construction, the term of correlated jumps should capture some contributions of asset returns from the common factor and from the idiosyncratic factor. Therefore, the decrease in the magnitudes of $\delta$ and $\xi$ compared with the benchmark model is likely and expected. Second, insurance companies experience the highest exposure to the common factor (0.72), followed by broker-dealers (0.59); others ha the lowest exposure (0.35).

**[Insert Table 3 Here]**

5.2. *Systemic risk measures: Preliminary analysis*

In this section we outline the stylized facts for three alternative systemic risk measures, based on both our model and the benchmark. The time series of the risk measures from 1996 to 2011 by sector appear in Panels A–C of Figure 3.

**[Insert Figure 3 Here]**

5.2.1. *DD*

Because the *DD* indicates how far a firm's asset value exceeds its default point for a given sector, it contrasts with conventional risk measures, such that a lower value of *DD* implies higher systemic risk for the sector. In Panel A of Figure 3, the tail dependence effects are of material importance if the red line appears below the blue line—as is the case in all sectors. The tail dependence effects reduce the distance to default during and prior to negative economic events, so *DD* is lower than $DD_{ben}$. For example, in the depositories sector, the tail dependence effects appeared;

(1) From the end of 1997 to mid-1999 (1997 Asian Crisis, 1998 LTCM debacle) with *DD* equal to 5.1 and $DD_{ben}$ equal to 5.9.[33]

(2) From September 2001 to 2003 (9/11 attack, end of dot.com bubble, credit market deterioration in 2002[34]) (*DD* 6.5, $DD_{ben}$ 9.9).

(3) From June 2006 (one year prior to the 2007 subprime loan crisis) to mid-2010 (2007–2010 financial crisis), with *DD* (4.5) versus $DD_{ben}$ (7.3).

(4) In the second half of 2011 (European debt crisis), with *DD* (2.4) and $DD_{ben}$ (3.6).

In the insurance sector, the effect arose in 2005–2009 (2005 automotive-downgrade credit crisis, 2007–2010 financial crisis), where *DD* was 10.7 and $DD_{ben}$ equaled 13.2. For others, the effect occurred at three moments: (1) from mid-1998 to mid-1999 (LTCM debacle) (*DD* 3.9, $DD_{ben}$ 5.5); (2) from September 2001 to September 2008 (9/11, end of dot.com bubble, credit market deterioration in 2002, low interest rates and high leverage among financial institutions during 2002–2004, 2007–2008 financial crisis) (*DD* 3.6, $DD_{ben}$ 7.1); and (3) during the second half of 2011 (European debt crisis) (*DD* 1.7, $DD_{ben}$ 3.8). In 2008–2009, the measures from both the full and benchmark models signaled that the others sector was very close to default.

---

[33] We compute average values of *DD* and $DD_{ben}$ and compare them for specific periods; for example, the average values from the end of 1997 to mid-1999 were 5.1 and 5.9, respectively.

[34] Huang *et al.* (2009) document that systemic risk exhibits substantial increases during 2002, due to the credit market deterioration.

5.2.2. *NoD*

Regarding the number of simultaneous defaults among the ten biggest financial institutions for each sector (Panel B, Figure 3), the tail dependence effect is significant for managers when the red line is above the blue line, that is, when *NoD* is larger than $NoD_{ben}$, which is mostly the case in our findings. That is, tail dependence effect increased significantly during the 2007–2010 financial crises across all four sectors. Before 2007, this effect was less noticeable than the *DD* measures were. For example, for depositories, we find this effect only in 1998 (*NoD* 0.35, $NoD_{ben}$ 0.04) and 2002 (*NoD* 0.29, $NoD_{ben}$ 0.02); for broker-dealers, it arose only between 1996 and 2003 (*NoD* 2.71, $NoD_{ben}$ 0.86). During the 2007-2010 crisis, *NoD* peaked, and the others sector emerged as the most risky, such that 9 of the 10 largest firms were expected to default. Depositories (8 of 10) and insurance companies (5 of 10) also exhibited substantial risk of default. Thus, risks in the financial industry increases through the channel of tail dependence in equity markets, especially in tough times.

5.2.3. *PIR*

Panel C of Figure 3 contains the time variation of *PIR*, or the ratio of the sector's price of insurance against financial distress to its aggregate asset value. The tail dependence effects are materially important for virtually all sectors, and the measure display especially strong effects of tail dependence during the financial crisis. Among broker-dealers for example, the *NoD* measure suggests similar levels of systemic risk for both the LTCM debacle and Lehman's bankruptcy (6 of 10 defaulting firms), but *PIR* signals greater systemic risk for the latter event (200) than the former (50). Empirical evidence also suggests that including tail dependence improves the model's ability to anticipate stressful periods. For example, in the others sector, *PIR* increased noticeably by October 2007, when Fannie Mae and Freddie Mac signaled their troubles due to the subprime crisis. In the broker-dealers and depositories sectors, *PIR* increased by March 2008, around the time of Bear Sterns's failure. However $PIR_{ben}$ did not show a clear upward trend until September 2008.

5.3. *Predictability*

A key criterion of the quality of a systemic risk indicator is its forecasting power. Therefore, we examine the lead-lag relationship between the full model–based measures and the benchmark-based ones. We use Granger causality whether our measures could forecast an index of financial distress. In particular, we used the St. Louis Fed Financial Stress Index (STLFSI), as proposed by Kliesen and Smith (2010).[35]

[35] The STLFSI is constructed by using 18 data series for different financial variables, including interest rates (effective federal funds rate, 2-year Treasury, 10-year Treasury, 30-year Treasury, Baa-rated

This index is publicly available and based on a principal component analysis of a broad range of financial prices and rates from different financial markets. Figure 4 shows the monthly time series of STLFSI from December 1996 to December 2011.[36] We find a local peak near the 1998 LTCM debacle, smooth increases between 2001 and 2002, increases after September 2007 (subprime crisis), a maximum level in September 2008 (Lehman bankruptcy), and two local peaks in mid-2010 and mid-2011 (acute stress periods in the Eurozone debt crisis).

**[Insert Figure 4 Here]**

### 5.3.1. *Granger causality test*

Because unit roots test offer conflicting results,[37] we rely on Granger causality (GC) tests in Table 4 for both levels (Panel A) and first differences (Panel B). For these tests, we use optimally chosen lags, corrected after controlling for heteroskedastic and correlated errors.[38]

Regarding the GC results for series in levels between the full model and

---

corporate, Merrill Lynch High-Yield Corporate Master II Index, and Merrill Lynch Asset-Backed Master BBB-rated), yield spreads (yield curve: 10-year Treasury minus 3-month Treasury, corporate Baa-rated bond minus 10-year Treasury, Merrill Lynch High-Yield Corporate Master II Index minus 10-year Treasury, 3-month London Interbank Offering Rate–Overnight Index Swap [LIBOR-OIS] spread, 3-month Treasury-Eurodollar [TED] spread, and 3-month commercial paper minus 3-month Treasury bill.), and other indicators (J.P. Morgan Emerging Markets Bond Index Plus, Chicago Board Options Exchange Market Volatility Index [VIX], Merrill Lynch Bond Market Volatility Index [1-month], 10-year nominal Treasury yield minus 10-year Treasury Inflation Protected Security yield, and Vanguard Financials Exchange-Traded Fund). Furthermore, the index is built by using principal component analysis to extract the factors responsible for the co-movement of a group of variables.

[36] We use monthly STLFSI, though the highest frequency is weekly, to match our data intervals.

[37] We also employed several unit root tests, including Augmented Dickey-Fuller, GLS Dickey-Fuller, and Perron (1997) with structural breaks in the mean, for the trends and both elements simultaneously. The detailed results are available on request.

[38] The optimal number of lags was chosen on the basis of Schwarz's Bayesian information criterion.

benchmark measures, the full model measures lead (usually by one or two months) benchmark-based ones in 10 of 12 cases; the remaining 2 cases exhibit bidirectional causality. That is, including the correlated jump factor improves the model's forecasting power in most cases over the benchmark.

The GC results for the comparison of the full model with the STLFSI in turn show that in 4 of 12 cases, the full model measures lead the STLFSI; in 2 cases, the STLFSI lead the full model; and 3 cases indicate bidirectional causality. Two broker-dealer sector measures (*DD*, *NoD*) and all the insurance sector systemic risk measures lead the STLFSI, by an average period of one month. Therefore, the measures in these two sectors are the most informative leading indicators. If *DD* and *NoD* measures increase in both sectors in a given month, a subsequent increase in the STLFSI index seems very likely indeed.

Using first difference data series (Panel B, Table 4), the full model–based measures lead (usually by one or two months) the benchmark-based ones in 8 of 12 cases; the reverse is true in 2 cases. These results generally agree with those we gathered from the series in levels. In the comparison of the full model measures and the STLFSI index, we find that the full model lead the STLFI in five cases, whereas the reverse occurs in 2 cases. That is, in agreement with series in levels, two broker-dealer systemic risk measures (*DD*, *NoD*) lead the STLFSI, as does one measure from the insurance sector

(*PIR*).

**[Insert Table 4 Here]**

Thus, the measures based on the full model contain more updated information than benchmark-based ones. Two measures related to broker-dealers (*DD* and *NoD*) and one measure in the insurance sector (*PIR*) provide leading information about the STLFSI index across all cases.[39]

5.3.2. *Predictive power*

To further compare the predictive ability of both models (benchmark and FM) in predicting STLFSI, we first run a predictive regression, including as explanatory variables the lagged terms of STLFSI and of the benchmark (pure common component model):

$$STLFSI_t = c + \sum_{s_1=1}^{k_1} \alpha_{s_1} STLFSI_{t-s_1} + \sum_{s_2=1}^{k_2} \beta_{s_2} Benchmark_{t-s_2} + \varepsilon_t. \quad (25)$$

Next we include the lagged terms of the FM factor (common plus extreme movement model) to determine the incremental predictive ability it may provide, using the following regression:

$$STLFSI_t = c + \sum_{s_1=1}^{k_1} \alpha_{s_1} STLFSI_{t-s_1} + \sum_{s_2=1}^{k_2} \beta_{s_2} Benchmark_{t-s_2} + \sum_{s_3=1}^{k_3} \omega_{s_3} FM_{t-s3} + \varepsilon_t. \quad (26)$$

where $k_1$, $k_2$, and $k_3$ are optimal lags selected according to the Bayesian information criterion.

[39] In addition to testing forecasting power over the whole sample period, we explore the full model measures could identify early warning signs of the 2007–2010 financial crisis better than the benchmark-based measures. We apply the Quandt-Andrews breakpoint test to date structural changes or break dates, identified by testing for structural changes in the coefficient of the autoregressive model with an order of 1 for the persistence test and of the constant term in the regressions for the level test. The changes should be primarily manifest in the leading indicators, then later in other variables. For the persistence test, the break dates identified by the full model measures across sectors all occur before July 2007 (the conventional crisis's starting point), and always lead the benchmark. The earliest two turning points happen for depositories and broker-dealers, in February 2006 and March 2006, respectively—that is, more than a year before July 2007. The full model measures also lead (coincide with) benchmark-based measures in 9 (1) of 12 cases. For the level test, the full model measures lead (coincide with) benchmark ones in 8 (2) cases. Overall the evidence supports the notion that considering tail dependence effects besides a common factor does provide more timely warning signals.

We use an *F-test* to determine if the difference in forecasting ability, as measured by $R^2$ values on the restricted model of Equation (25) and the unrestricted model of Equation (26), differs significantly from zero. Formally, the *F-statistic* is computed as

$$F\text{-}statistic = [(R^2_{_eq.(2)} - R^2_{_eq.(1)}) / (k_1+k_2+k_3 - (k_1+k_2))\ ] / [\ (1- R^2_{_eq.(2)}) / (N - (k_1+k_2+k_3) - 1)\ ], \quad (27)$$

where $N$ is the sample size, and the degrees of freedom are computed as $v_1= (k_1+k_2+k_3 - (k_1+k_2))$ and $v_2= (N - (k_1+k_2+k_3) - 1)$. Table 5 reports results and reveals cases where $R^2$ is higher in Equation (26) than in Equation (25), using bold font.

For the data series in levels (Panel A of Table 5), in 8 of 12 cases the FM models offer additional explanatory power, as indicated by the higher $R^2$ for Equation (26) than Equation (25). This additional explanatory power is particularly significant in five cases: *DD* on depositories, *DD* on insurance companies, *NoD* on others, *PIR* on broker-dealers, and *PIR* on insurance companies. For data series in first differences (Panel B of Table 5), in 10 of 12 cases, FM models have some additional explanatory power, especially notable in five cases: *DD* on insurance companies, *NoD* on depositories, *NoD* and *PIR* on broker-dealers, and *PIR* on insurance companies. The evidence thus suggests that FM-based measures have extra predictive power in comparison with the benchmark model, especially in the case of *DD* on insurance companies and *PIR* on broker-dealers and insurance companies.

**[Insert Table 5 Here]**

## 6. Conclusion

Growing evidence suggests that systemic risk results from at least two driving forces: the common factor exposure to market-wide shocks and the tail dependence effects that arise from links among extreme stock returns. Modeling the relative importance of these two factors is critical; we seek to contribute to this literature stream by proposing a new structural-form model that includes both factors. For our framework, the common factor component is based on correlations of a financial institution's individual stock returns with an aggregate common factor, and we proxy for tail dependence effects with a correlated jumps factor. The empirical implications of our model tests are consistent with extant evidence; in particular, they suggest that simultaneous extreme negative movements across large financial institutions are stronger in bear markets than in bull markets.

With an empirical application based on stock market data for four sectors of the U.S. financial industry during 1996–2011, we demonstrate that ignoring the effect of tail dependence will lead to underestimates of the level of systemic risk. By accounting for tail dependence effects, we gain extra forecasting power, compared with a benchmark model. Not all sectors provide equally valuable systemic risk indicators though. Rather, two measures (*DD*, *NoD*) in the broker-dealer sector and one measure (*PIR*) from the insurance sector systematically lead the St. Louis Fed Financial Stress Index (STLFSI).

Looking forward, a comparison of our measures with other measures based on alternative asset markets would offer an interesting topic for further investigation. The application of our measures for asset pricing, hedging strategies, portfolio diversification, and risk management purposes represent other natural directions for further research.

# Appendices

## Appendix A

We apply the theorem of the law of total variance,

$$Var(Y) = E_X\left[Var(Y \mid X)\right] + Var_X\left[E(Y \mid X)\right]. \quad (A.1)$$

In our case, we have,

$$Y \equiv Q_j N(\Delta t) = \sum_{k=1}^{N(\Delta t)} Q_j^{(k)}(\Delta t), \text{ and } X = N(\Delta t). \quad (A.2)$$

Therefore,

$$\begin{aligned} Var^P\left(Q_j N(\Delta t)\right) &= E_N\left[b_j^2 N(\Delta t)\right] + Var_N\left[N(\Delta t) a_j\right] \\ &= b_j^2 \lambda + a_j^2 \lambda \\ &= \left[a_j^2 + b_j^2\right]\lambda \end{aligned} \quad . (A.3)$$

Finally, we assume that all random variables appear in the asset–log return process described by Equation (1) are independent and derive the variance of asset returns as follows:

$$Var\left(v_{j,t} \mid \varphi_{t-1}\right) \equiv \sigma_{j,t}^2 = \delta_j^2 h_t + \xi_j + \lambda \hat{b}_j^2, \quad (A.4)$$

where $\hat{b}_j^2 = a_j^2 + b_j^2$.

## Appendix B

Because $Q_k$ are normally i.i.d. random variables, distributed independently of $N(T)$, by iterated expectations, we know

$$\begin{aligned} E\left[e^{\phi Q_j N(T)}\right] &= E\left[e^{\phi \sum_{k=1}^{N(T)} Q_{j,k}}\right] = E\left[\prod_{k=1}^{N(T)} e^{\phi Q_{j,k}}\right] = E_N\left[E_{Q_j|N}\left[\prod_{k=1}^{N(T)} e^{\phi Q_{j,k}} \mid N(T)\right]\right] \\ &= \sum_{i=0}^{\infty} p_i(\lambda T) E\left[\prod_{k=1}^{i} e^{\phi Q_{j,k}}\right] = \sum_{i=0}^{\infty} p_i(\lambda T) \prod_{k=1}^{i} E\left[e^{\phi Q_{j,k}}\right] \\ &= \sum_{i=0}^{\infty} \frac{e^{-\lambda T}(\lambda T)^i}{i!}\left(e^{\left(a_j\phi + \frac{1}{2}b_j^2\phi^2\right)}\right)^i = e^{-\lambda T}\sum_{i=0}^{\infty}\frac{1}{i!}\left(\lambda T e^{\left(a_j\phi+\frac{1}{2}b_j^2\phi^2\right)}\right)^i \\ &= e^{-\lambda T} e^{\lambda T \exp\left(a_j\phi + \frac{1}{2}b_j^2\phi^2\right)} = e^{\lambda T\left(\exp\left(a_j\phi+\frac{1}{2}b_j^2\phi^2\right)-1\right)} \end{aligned} \quad . (B.1)$$

## Appendix C

| Type | Company Name | Start Date | End Date | Number of Observations | Size (millions) | LVG |
|---|---|---|---|---|---|---|
| Depositories | 'BANK OF AMERICA CORP' | 199601 | 201112 | 181 | 13.556 | 9.194 |
| Depositories | 'BANK OF NEW YORK MELLON CORP' | 200310 | 201112 | 55 | 12.037 | 5.709 |
| Depositories | 'BANK ONE CORP' | 199601 | 200406 | 91 | 12.101 | 5.949 |
| Depositories | 'BANKAMERICA CORP-OLD' | 199601 | 199809 | 22 | 12.391 | 7.568 |
| Depositories | 'BANKERS TRUST CORP' | 199601 | 199905 | 30 | 11.693 | 16.393 |
| Depositories | 'BB&T CORP' | 200401 | 201112 | 72 | 11.745 | 7.345 |
| Depositories | 'CITICORP' | 199601 | 199809 | 22 | 12.510 | 6.568 |
| Depositories | 'FIFTH THIRD BANCORP' | 200308 | 200708 | 20 | 11.473 | 4.391 |
| Depositories | 'FIRST CHICAGO NBD CORP' | 199601 | 199809 | 22 | 11.639 | 7.736 |
| Depositories | 'FLEETBOSTON FINANCIAL CORP' | 199604 | 200403 | 69 | 11.846 | 5.777 |
| Depositories | 'GOLDEN WEST FINANCIAL CORP' | 200501 | 200609 | 10 | 11.585 | 6.224 |
| Depositories | 'JPMORGAN CHASE & CO' | 199601 | 201112 | 181 | 13.565 | 10.194 |
| Depositories | 'KEYCORP' | 199712 | 200408 | 27 | 11.294 | 7.799 |
| Depositories | 'MORGAN (J P) & CO' | 199601 | 200012 | 49 | 12.383 | 12.952 |
| Depositories | 'NATIONAL CITY CORP' | 199807 | 200812 | 109 | 11.614 | 6.439 |
| Depositories | 'PNC FINANCIAL SVCS GROUP INC' | 199711 | 201112 | 47 | 12.049 | 8.966 |
| Depositories | 'REGIONS FINANCIAL CORP' | 200704 | 201112 | 46 | 11.856 | 17.872 |
| Depositories | 'STATE STREET CORP' | 200305 | 201112 | 50 | 11.875 | 8.214 |
| Depositories | 'SUNTRUST BANKS INC' | 199904 | 201112 | 142 | 11.829 | 9.124 |
| Depositories | 'U S BANCORP' | 200107 | 201112 | 115 | 12.260 | 5.103 |
| Depositories | 'U S BANCORP/DE-OLD' | 200010 | 200205 | 9 | 11.369 | 4.466 |
| Depositories | 'WACHOVIA CORP' | 199601 | 200812 | 145 | 12.586 | 6.926 |
| Depositories | 'WASHINGTON MUTUAL INC' | 199801 | 200808 | 117 | 12.334 | 8.449 |
| Depositories | 'WELLS FARGO & CO -OLD' | 199610 | 199810 | 14 | 11.572 | 4.678 |
| Depositories | 'WELLS FARGO & CO' | 199601 | 201112 | 165 | 12.771 | 5.456 |
| Broker-Dealers | 'AMERIPRISE FINANCIAL INC' | 200604 | 201112 | 58 | 11.560 | 11.357 |
| Broker-Dealers | 'AXA FINANCIAL INC' | 199601 | 200012 | 49 | 11.872 | 18.052 |
| Broker-Dealers | 'BEAR STEARNS COMPANIES INC' | 199601 | 200805 | 138 | 12.091 | 27.973 |
| Broker-Dealers | 'BLACKROCK INC' | 200701 | 201112 | 49 | 10.378 | 3.734 |
| Broker-Dealers | 'CITIGROUP GLOBAL MKTS HLDGS' | 199601 | 199710 | 11 | 12.095 | 41.126 |
| Broker-Dealers | 'CREDIT SUISSE USA INC' | 199604 | 200010 | 44 | 11.085 | 18.551 |
| Broker-Dealers | 'DAIN RAUSCHER CORP' | 199601 | 199702 | 3 | 7.725 | 8.450 |
| Broker-Dealers | 'E TRADE FINANCIAL CORP' | 200002 | 201112 | 132 | 10.338 | 14.236 |
| Broker-Dealers | 'EDWARDS (A G) INC' | 199601 | 200207 | 49 | 8.304 | 1.880 |
| Broker-Dealers | 'FRANKLIN RESOURCES INC' | 199601 | 200702 | 56 | 8.805 | 1.194 |
| Broker-Dealers | 'GOLDMAN SACHS GROUP INC' | 199909 | 201112 | 137 | 13.195 | 11.249 |
| Broker-Dealers | 'INTERACTIVE BROKERS GROUP' | 200710 | 201112 | 40 | 10.300 | 36.641 |
| Broker-Dealers | 'JEFFERIES GROUP INC' | 200107 | 201112 | 97 | 9.717 | 7.315 |
| Broker-Dealers | 'LEGG MASON INC' | 200104 | 200311 | 19 | 8.614 | 2.492 |
| Broker-Dealers | 'LEHMAN BROTHERS HOLDINGS INC' | 199601 | 200808 | 141 | 12.393 | 22.165 |
| Broker-Dealers | 'MERRILL LYNCH & CO INC' | 199601 | 200812 | 145 | 12.947 | 12.367 |
| Broker-Dealers | 'MORGAN STANLEY' | 199601 | 201112 | 181 | 13.173 | 14.750 |
| Broker-Dealers | 'PAINE WEBBER GROUP' | 199601 | 200010 | 47 | 10.921 | 16.013 |
| Broker-Dealers | 'QUICK & REILLY GROUP INC' | 199603 | 199801 | 12 | 8.117 | 5.102 |
| Broker-Dealers | 'RAYMOND JAMES FINANCIAL CORP' | 199703 | 201112 | 102 | 8.975 | 5.428 |
| Broker-Dealers | 'SCHWAB (CHARLES) CORP' | 199604 | 201112 | 178 | 10.452 | 3.052 |
| Broker-Dealers | 'SWS GROUP INC' | 199707 | 200202 | 15 | 8.316 | 14.279 |
| Broker-Dealers | 'TD AMERITRADE HOLDING CORP' | 200301 | 201112 | 94 | 9.667 | 2.903 |
| Broker-Dealers | 'TD WATERHOUSE GROUP INC' | 199911 | 200110 | 13 | 9.238 | 2.259 |
| Insurance Companies | 'AETNA INC' | 199601 | 200011 | 48 | 11.476 | 9.204 |
| Insurance Companies | 'AFLAC INC' | 200904 | 201112 | 22 | 11.314 | 4.714 |
| Insurance Companies | 'ALLSTATE CORP' | 199607 | 201112 | 175 | 11.666 | 5.097 |
| Insurance Companies | 'AMERICAN GENERAL CORP' | 199601 | 200107 | 56 | 11.336 | 7.109 |
| Insurance Companies | 'AMERICAN INTERNATIONAL GROUP' | 199601 | 201112 | 181 | 13.028 | 36.671 |
| Insurance Companies | 'CIGNA CORP' | 199601 | 200508 | 105 | 11.479 | 9.051 |
| Insurance Companies | 'CNA FINANCIAL CORP' | 199601 | 200302 | 63 | 11.049 | 9.223 |
| Insurance Companies | 'CNO FINANCIAL GROUP INC' | 200001 | 200108 | 7 | 10.826 | 15.516 |
| Insurance Companies | 'GENERAL RE CORP' | 199601 | 199705 | 6 | 10.476 | 3.369 |
| Insurance Companies | 'GENWORTH FINANCIAL INC' | 200410 | 201112 | 76 | 11.587 | 23.942 |
| Insurance Companies | 'HANCOCK JOHN FINL SVCS INC' | 200007 | 200403 | 34 | 11.399 | 9.408 |
| Insurance Companies | 'HARTFORD FINANCIAL SERVICES' | 199604 | 201112 | 178 | 12.231 | 19.248 |
| Insurance Companies | 'HARTFORD LIFE INC -CL A' | 199710 | 200005 | 21 | 11.566 | 83.438 |
| Insurance Companies | 'LINCOLN NATIONAL CORP' | 199601 | 201112 | 181 | 11.627 | 15.279 |
| Insurance Companies | 'LOEWS CORP' | 199601 | 201002 | 106 | 11.185 | 7.264 |
| Insurance Companies | 'METLIFE INC' | 200010 | 201112 | 124 | 12.878 | 13.855 |
| Insurance Companies | 'NATIONWIDE FINL SVCS -CL A' | 199807 | 200812 | 115 | 11.490 | 68.231 |
| Insurance Companies | 'PRINCIPAL FINANCIAL GRP INC' | 200204 | 201112 | 106 | 11.706 | 13.231 |
| Insurance Companies | 'PROVIDIAN CORP' | 199601 | 199702 | 3 | 10.179 | 6.585 |
| Insurance Companies | 'PRUDENTIAL FINANCIAL INC' | 200204 | 201112 | 106 | 12.903 | 17.108 |
| Insurance Companies | 'TRANSAMERICA CORP' | 199601 | 199805 | 12 | 10.782 | 9.311 |
| Insurance Companies | 'TRAVELERS COS INC' | 199610 | 201112 | 85 | 11.542 | 5.642 |
| others | 'AMERICAN EXPRESS CO' | 199601 | 201112 | 181 | 11.817 | 3.668 |
| others | 'ANNALY CAPITAL MANAGEMENT' | 200207 | 201112 | 55 | 10.721 | 7.537 |
| others | 'APARTMENT INVST & MGMT CO' | 200204 | 200308 | 6 | 9.102 | 2.843 |
| others | 'ASSOCIATES FIRST CAP -CL A' | 199610 | 200011 | 39 | 10.991 | 6.809 |
| others | 'BENEFICIAL CORP' | 199601 | 199806 | 19 | 9.670 | 5.548 |
| others | 'CAPITAL ONE FINANCIAL CORP' | 200007 | 201112 | 127 | 11.083 | 5.916 |
| others | 'CAPSTEAD MORTGAGE CORP' | 199601 | 200205 | 40 | 9.256 | 18.884 |
| others | 'CIT GROUP INC' | 200301 | 201112 | 81 | 10.998 | 10.818 |
| others | 'CIT GROUP INC-OLD' | 199804 | 200105 | 27 | 10.143 | 11.133 |
| others | 'CITIGROUP INC' | 199601 | 201112 | 181 | 13.683 | 13.404 |
| others | 'CME GROUP INC' | 200901 | 201112 | 25 | 10.546 | 1.644 |
| others | 'COUNTRYWIDE FINANCIAL CORP' | 199601 | 200806 | 139 | 10.463 | 5.476 |
| others | 'DEAN WITTER DISCOVER & CO' | 199601 | 199705 | 6 | 10.476 | 4.333 |
| others | 'DISCOVER FINANCIAL SVCS INC' | 200712 | 201112 | 38 | 10.677 | 6.766 |
| others | 'FANNIE MAE' | 199601 | 201006 | 163 | 13.400 | 84.908 |
| others | 'FEDERAL HOME LOAN MORTG CORP' | 199604 | 201006 | 160 | 13.097 | 125.524 |
| others | 'FINOVA GROUP INC' | 199601 | 200201 | 44 | 9.147 | 23.722 |
| others | 'FIRST USA INC' | 199601 | 199705 | 6 | 8.864 | 3.004 |
| others | 'GENERAL GROWTH PPTYS INC' | 200504 | 201111 | 11 | 10.223 | 5.368 |
| others | 'HELLER FINANCIAL INC' | 199810 | 200109 | 25 | 9.679 | 15.190 |
| others | 'HOST HOTELS & RESORTS INC' | 200107 | 200305 | 12 | 9.025 | 3.650 |
| others | 'HSBC FINANCE CORP' | 199601 | 200302 | 75 | 10.739 | 3.684 |
| others | 'IMPAC MORTGAGE HOLDINGS INC' | 200603 | 200705 | 4 | 10.231 | 37.595 |
| others | 'INTERCONTINENTALEXCHANGE INC' | 201101 | 201112 | 1 | 10.248 | 3.902 |
| others | 'MF GLOBAL HOLDINGS LTD' | 200801 | 201109 | 34 | 10.828 | 65.259 |
| others | 'NELNET INC' | 200708 | 201112 | 17 | 10.246 | 43.709 |
| others | 'NEW CENTURY FINANCIAL CORP' | 200601 | 200701 | 2 | 10.278 | 13.369 |
| others | 'SIMON PROPERTY GROUP INC' | 199901 | 201102 | 58 | 9.564 | 3.249 |
| others | 'SLM CORP' | 200210 | 201112 | 100 | 11.578 | 16.639 |
| others | 'STUDENT LOAN CORP' | 199607 | 201012 | 83 | 9.841 | 11.464 |
| others | 'THORNBURG MORTGAGE INC' | 200310 | 200811 | 51 | 10.414 | 15.003 |

## References

Acharya, V. V., Pedersen, L. H., Philippon, T. and Richardson, M., ‘Measuring systemic risk’, *Technical report* (Department of Finance, NYU, 2010).

Adrian, T. and Brunnermeier, M. K., ‘CoVaR’, *Working paper* (Federal Reserve Bank of New York Staff Reports, No. 348, 2010).

Altman, Ed., and Rijken, H., ‘Toward a Bottom-up Approach to Assessing Sovereign Default Risk’, *Journal of Applied Corporate Finance*, Vol. 23, 2011, pp. 20–31.

Aramonte, S., Rosen, S. and Schindler, J.W. ‘Assessing and Combining Financial Conditions Indexes’, Finance and Economics Discussion Series of the Board of Governors of the Federal Reserve; 30 May 2013, No. 2013–39.

Bae, K-H, Karolyi, G. A. and Stulz, R. M., ‘A new approach to measuring financial Contagion’, *Review of Financial Studies*, Vol. 16, 2003, pp. 717–764.

Bates, D. S., ‘The Crash of '87: was it expected? the evidence from options markets’, *Journal of Finance*, Vol. 46, 1991, pp. 1009–1044.

Brownlees, C. T. and Engle, R., ‘Volatility, correlation and tails for systemic risk measurement’, *Working Paper* (Stern School of Business, New York University, 2011).

Bisias, D., Flood, M., Lo, A. W. and Valavanis, S., ‘A survey of systemic risk analytics’, *Working Paper* (Office of Financial Research, 2012)

Carlson, M. A., King, T. B. and Lewis, K. F., ‘Distress in the financial sector and economic activity’, *The BE Journal of Economic Analysis Policy*, Vol. 11, 2011, pp. 1–29.

Cummins, J. D. and Weiss, M.A., ‘Systemic risk and the U.S. insurance sector’, *Working Pap*er (Temple University, 2010)

Das S. and Uppal, R., ‘Systemic Risk and International Portfolio Choice’, *Journal of Finance*, Vol. 59, 2004, pp. 2809–2834.

Das, S. R., Duffie, D., Kapadia, N. and Saita, L., ‘Common failings: how corporate defaults are correlated’, *Journal of Finance*, Vol. 62, 2007, pp. 93–117.

Duan, J.-C., ‘Maximum likelihood estimation using the price data of the derivative contract’, *Mathematical Finance*, Vol. 4, 1994, pp. 155–167.

Duan, J.-C., ‘The GARCH option pricing model’, *Mathematical Finance*, Vol. 5, 1995, pp. 13–32.

Duan, J.-C., ‘Correction: maximum likelihood estimation using the price data of the derivative contract’, *Mathematical Finance*, Vol. 10, 2000, pp. 461–462.

Duffie, D., A. Eckner, G. Horel, and L. Saita., “Frailty Correlated Default”, Journal of Finance, Vol. 64, 2009, pp. 2089–2123.

Elsinger, H., Lehar, A. and Summer, M., ‘Risk assessment for banking systems’, *Management Science*, Vol. 52, 2006a, pp. 1301–1314.

Elsinger, H., Lehar, A. and Summer, M., ‘Using market information for banking system risk assessment’, *International Journal of Central Bank*, Vol. 2, 2006b, pp. 137–165.

Engle, R., ‘Dynamic conditional correlation—a simple class of multivariate GARCH models’, *Journal of Business and Economic Statistics*, Vol. 20, 2002, pp. 339–350.

Forte, S. and Peña, J. I., ‘Credit spreads: An empirical analysis on the informational content of stocks, bonds, and CDS’, *Journal of Banking and Finance*, Vol. 33, 2009, pp. 2013–2025.

Giesecke, K., ‘Correlated default with incomplete information’, *Journal of Banking and Finance*, Vol. 28, 2004, pp. 1521–1545.

Giesecke, K. and Goldberg, L. R., ‘Sequential defaults and incomplete information’, *Journal of Risk*, Vol. 7, 2004, pp. 1–26.

Giesecke, K. and Kim, B., ‘Systemic risk: what defaults are telling us’, *Management* Science, Vol. 57, 2011, pp. 1387–1405.

Gropp, R., Lo Duca, M., Vesala, J., ‘Cross-border bank contagion in Europe’, *International Journal of Central Banking*, Vol. 5, 2009, pp. 97–139.

Heston, S. and Nandi, S., ‘A closed-form GARCH option valuation model’, *Review of Financial Studies*, Vol. 13, 2000, pp. 585–625.

Huang, X., Zhou, H. and Zhu, H., ‘A framework for assessing the systemic risk of major financial institutions’, *Journal of Banking Finance*, Vol. 33, 2009, pp. 2036–2049.

IMF, ‘Global financial stability report: Grappling with crisis legacies’. *Technical report* (International Monetary Fund, Washington, 2011).

Jobst, A., ‘Measuring systemic risk-adjusted liquidity (SRL) - A Model Approach’, *Working Paper* (IMF, 2012).

Jokipii, T. and Monnin, P., ‘The impact of banking sector stability on the real economy’, *Journal of International Money and Finance*, Vol. 32, 2013, pp. 1–16.

Jorion, P. and Zhang, G., ‘Credit contagion from counterparty risk’, *Journal of Finance*,

Vol. 64, 2009, pp. 2053–2087.
Kliesen, K. L. and Smith, D. C., ‘Measuring financial market stress’, *Federal Reserve Bank of St. Louis Economic Synopses*, No. 2, 2010.
Krainer, J., and Lopez, J. A., ‘Incorporating equity market information into supervisory monitoring models’, *Journal of Money, Credit, and Banking*, Vol. 36, 2004, pp. 1043–1067.
Lehar, A., ‘Measuring systemic risk: A risk management approach’, *Journal of Banking and Finance*, Vol. 29, 2005, pp. 2577–2603.
Liu, S., Qi, H., Shi, J. and Xie, Y. A., ‘Inferring Default Correlation from Equity Return Correlation’, *European Financial Management*, 2013, forthcoming.
Merton, R., ‘On the pricing of corporate debt: the risk structure of interest rates’, *Journal of Finance*, Vol. 29, 1974, pp. 449–470.
Norden, L. and Weber, M., ‘The co-movement of credit default swap, bond and stock markets: an empirical analysis’, *European Financial Management*, Vol. 15, 2009, pp. 529–562.
Pais, A. and Stork, P. A., ‘Bank size and systemic risk’, *European Financial Management*, Vol. 19, 2013, pp. 429–451.
Patro, D. K., Qi, M. and Sun, X., ‘A simple indicator of systemic risk’, *Journal of Financial Stability*, Vol. 9, 2013, pp. 105–116.
Perron, P., ‘Further evidence on breaking trend functions in macroeconomic variables’, *Journal of Econometrics*, Vol. 80, 1997, pp. 355–385.
Rajan, R. G., ‘Has Finance Made the World Riskier?’ *European Financial Management*, Vol. 12, 2006, pp. 499–533.
Rodriguez Moreno, M. and Peña, J.I., ‘Systemic Risk Measures: The simpler the better?’, *Journal of Banking and Finance*, Vol. 37, 2013, pp. 1817–1831.
Ronn, E. I. and Verma, A. K., ‘Pricing risk-adjusted deposit insurance: an option-based model’, *Journal of Finance*, Vol. 41, 1986, pp. 871–895.
Saldias, M. ‘A market-based approach to sector risk determinants and transmission in the euro area’, *Journal of Banking and Finance*, 2013, forthcoming
Silvennoinen, A. and T. Teräsvirta (2009) “Multivariate GARCH models”, in T. G. Andersen, R. A. Davis, J.-P. Kreiss and T. Mikosch, eds. Handbook of Financial Time Series. New York: Springer.
Suh, S., ‘Measuring systemic risk: A factor-augmented correlated default approach’, *Journal of Financial Intermediation*, Vol. 21, 2012, pp. 341–358.
Vassalou, M. and Xing, Y., ‘Default risk in equity returns’, *Journal of Finance*, Vol. 59, 2004, pp. 831–868.
Zhang, B. Y., Zhou, H. and Zhu, H., ‘Explaining credit default swap spreads with equity volatility and jump risks of individual firms’, *Review of Financial Studies*, Vol. 22, 2009, pp. 5099–5131.
Zhou, C., ‘An analysis of default correlation and multiple defaults’, *Review of Financial Studies*, Vol. 14, 2001a, pp. 555–576.
Zhou, C., ‘The term structure of credit spreads with jump risk’, *Journal of Banking and Finance*, Vol. 25, 2001b, pp. 2015–2040.

**Table 1. Summary statistics.**

This table reports summary statistics for several risk measures for each sector of the financial industry, from December 1996 to December 2011, for a total of 181 monthly observations. *SIZE* (in millions) is the logarithm of aggregated total assets for the ten biggest firms in each sector. The *LVG* is the quasi-market value of assets divided by the market value of equity, with weighted averages based on the values of market equity. The *RET* is annual returns. *DD*, *NoD*, and *PIR* (scaled by multiplying them by $10^6$) are systemic risk measures. The subindex "ben" identifies measures computed from the benchmark model (without correlated jump terms).

| Sector | Statistics | *SIZE* | *LVG* | *RET* | *DD* | *NoD* | *PIR* | $DD_{ben}$ | $NoD_{ben}$ | $PIR_{ben}$ |
|---|---|---|---|---|---|---|---|---|---|---|
| Depositories | Min | 14.380 | 4.890 | -0.530 | -0.240 | 0.000 | 0.000 | 1.120 | 0.000 | 0.000 |
| | Max | 15.800 | 34.040 | 1.760 | 16.230 | 8.030 | 69.255 | 16.730 | 3.860 | 9.475 |
| | Mean | 15.140 | 7.940 | 0.090 | 6.310 | 0.790 | 4.503 | 8.000 | 0.260 | 0.343 |
| | Median | 15.070 | 6.910 | 0.070 | 5.900 | 0.010 | 0.020 | 7.430 | 0.000 | 0.000 |
| | Std | 0.440 | 3.400 | 0.260 | 3.930 | 1.720 | 11.765 | 4.030 | 0.650 | 1.123 |
| Broker-Dealers | Min | 13.760 | 6.240 | -0.540 | -0.660 | 0.000 | 0.005 | 0.020 | 0.000 | 0.000 |
| | Max | 15.350 | 40.050 | 1.380 | 8.440 | 6.110 | 208.445 | 16.700 | 5.250 | 29.956 |
| | Mean | 14.540 | 12.260 | 0.210 | 2.590 | 2.260 | 22.221 | 6.010 | 0.960 | 2.614 |
| | Median | 14.470 | 10.960 | 0.160 | 1.980 | 2.140 | 6.334 | 5.140 | 0.790 | 0.693 |
| | Std | 0.390 | 4.800 | 0.390 | 2.110 | 1.820 | 39.363 | 3.800 | 1.060 | 4.785 |
| Insurance | Min | 13.450 | 4.140 | -0.540 | 0.910 | 0.000 | 0.000 | 0.390 | 0.000 | 0.000 |
| | Max | 15.020 | 82.960 | 2.200 | 21.230 | 5.390 | 41.335 | 26.290 | 3.220 | 22.541 |
| | Mean | 14.460 | 11.930 | 0.120 | 10.990 | 0.360 | 3.757 | 12.180 | 0.230 | 1.421 |
| | Median | 14.600 | 7.620 | 0.120 | 10.870 | 0.000 | 0.000 | 11.880 | 0.000 | 0.000 |
| | Std | 0.470 | 14.160 | 0.340 | 4.570 | 1.010 | 10.084 | 5.380 | 0.580 | 4.281 |
| Others | Min | 13.450 | 4.310 | -0.670 | -0.780 | 0.000 | 0.001 | -1.970 | 0.000 | 0.000 |
| | Max | 15.390 | 173.300 | 1.900 | 7.390 | 8.900 | 288.020 | 11.260 | 6.770 | 211.190 |
| | Mean | 14.810 | 11.220 | 0.110 | 3.450 | 2.380 | 39.902 | 5.510 | 1.420 | 15.074 |
| | Median | 14.890 | 7.690 | 0.110 | 3.750 | 1.450 | 3.699 | 6.040 | 1.000 | 0.919 |
| | Std | 0.490 | 16.920 | 0.340 | 2.170 | 2.400 | 78.679 | 3.150 | 1.730 | 39.273 |

**Table 2. Estimation results: correlated jumps.**

Columns 1, 3, and 5 provide the average values of $\lambda$, *mu_coj*, and *std_coj* for the pre-crisis, crisis, and post-crisis periods, respectively. The pre-crisis period runs from July 2005 to June 2007, the crisis period from July 2007 to June 2009, and the post-crisis period is from July 2009 to June 2011. Each period consists of 24 observations. Columns 2 and 4 report the results of the independent samples t-test, for which the null hypothesis is that the means of the two groups are equal. For each parameter, Column 2 (4) reports the difference in the average values for crisis and pre-crisis (post-crisis) periods, with the *p*-values in brackets. Panel A refers to depositories, Panel B to broker-dealers, Panel C to insurance companies, and Panel D to others sectors.

| | Pre-Crisis<br>(1) | Crisis versus Pre-Crisis<br>(2) | Crisis<br>(3) | Crisis versus Post-Crisis<br>(4) | Post-Crisis<br>(5) |
|---|---|---|---|---|---|
| Panel A: Depositories | | | | | |
| $\lambda$ | 0.0394 | 0.0964 *** | 0.1358 | 0.0828 *** | 0.0530 |
| | | (<0.0001) | | (0.0015) | |
| *mu_coj* | 0.0049 | -0.0135 *** | -0.0086 | -0.0078 * | -0.0008 |
| | | (0.0007) | | (0.1039) | |
| *std_coj* | 0.0178 | 0.0653 *** | 0.0831 | 0.0128 | 0.0702 |
| | | (<0.0001) | | (0.3478) | |
| Panel B: Broker-Dealers | | | | | |
| $\lambda$ | 0.0347 | 0.0258 ** | 0.0606 | 0.0074 | 0.0532 |
| | | (0.0244) | | (0.5427) | |
| *mu_coj* | 0.0030 | -0.0162 ** | -0.0132 | -0.0101 | -0.0031 |
| | | (0.0476) | | (0.2235) | |
| *std_coj* | 0.0318 | 0.0535 *** | 0.0853 | 0.0256 *** | 0.0597 |
| | | (<0.0001) | | (0.0054) | |
| Panel C: Insurance Companies | | | | | |
| $\lambda$ | 0.0713 | 0.0561 *** | 0.1274 | 0.0626 *** | 0.0648 |
| | | (0.0030) | | (0.0011) | |
| *mu_coj* | 0.0013 | -0.0041 *** | -0.0028 | -0.0024 ** | -0.0004 |
| | | (<0.0001) | | (0.0177) | |
| *std_coj* | 0.0206 | 0.0634 *** | 0.0840 | -0.0002 | 0.0842 |
| | | (<0.0001) | | (0.9896) | |
| Panel C: Others | | | | | |
| $\lambda$ | 0.1003 | 0.0963 *** | 0.1967 | 0.1026 *** | 0.0941 |
| | | (0.0015) | | (0.0031) | |
| *mu_coj* | -0.0008 | -0.0222 ** | -0.0230 | -0.0234 ** | 0.0004 |
| | | (0.0187) | | (0.0160) | |
| *std_coj* | 0.0273 | 0.1029 *** | 0.1302 | 0.0483 *** | 0.0820 |
| | | (<0.0001) | | (0.0081) | |

**Table 3. Estimation results: structural-form parameters.**

This table reports the average values of the estimated parameters from our structural-form model and the benchmark model (notated "ben"). The sample period spans December 1996 to December 2011, with 181 observations. The reported numbers in $\mu$, $\mu_{_ben}$, $\xi$, and $\xi_{_ben}$ are 10,000 times the raw values. The "diff." indicates stands for the testing results of independent samples t-tests, for which the null hypothesis is that the means of the two groups are equal for each pair of parameters. Columns 3, 6, and 9 report the differences of the average values of the estimated parameters, along with *p*-values in brackets.

| Sector | $\mu$ (1) | $\mu_{_ben}$ (2) | *diff.* (3) | $\delta$ (4) | $\delta_{_ben}$ (5) | *diff.* (6) | $\xi$ (7) | $\xi_{_ben}$ (8) | *diff.* (9) |
|---|---|---|---|---|---|---|---|---|---|
| Depositories | 0.9996 | 0.9247 | 0.0749 | 0.4486 | 0.5193 | -0.0707 *** | 0.2714 | 0.4960 | -0.2246 *** |
| | | | (0.8814) | | | (<0.0001) | | | (<0.0001) |
| Broker-Dealers | 2.8938 | 3.1188 | -0.2250 | 0.5945 | 0.6326 | -0.0381 *** | 0.5957 | 1.0429 | -0.4472 *** |
| | | | (0.6611) | | | (0.0012) | | | (<0.0001) |
| Insurance Companies | -0.9960 | -0.1626 | -0.8334 | 0.7224 | 0.7842 | -0.0618 *** | 0.5299 | 1.5100 | -0.9800 *** |
| | | | (0.2683) | | | (0.0003) | | | (<0.0001) |
| Others | 2.7060 | 3.2008 | -0.4948 | 0.3534 | 0.4157 | -0.0622 *** | 0.1897 | 0.8126 | -0.6229 *** |
| | | | (0.2911) | | | (<0.0001) | | | (<0.0001) |

**Table 4. Granger causality tests, 1996–2011.**

This table reports the results of Granger causality tests for full-model (FM) systemic risk indicators compared with benchmark-based ones and with the public financial stress index STLFSI. The sample contains 181 monthly observations from December 1996 to December 2011. Panel A refers to levels and Panel B includes the results for first differences. The tests for each risk indicator apply across industries. Columns 1 and 2 (3 and 4) indicate if the systemic risk indicators Granger cause the benchmark-based measures (STLFSI), and their reverse direction. The Granger causality tests' lag-lengths are selected according to the Schwarz criterion; heteroskedastic and correlated errors corrected. For each test, the *p*-values appear in brackets, and the lag-length of VAR is reported for statistically significant cases.
***, **, and * are significant at 1%, 5%, and 10% levels, respectively.

| | | FM leads Benchmark (1) | Benchmark leads FM (2) | FM leads STLFSI (3) | STLFSI leads FM (4) |
|---|---|---|---|---|---|
| Measure | Sector | *P*-value | *P*-value | *P*-value | *P*-value |
| Panel A: levels | | | | | |
| *DD* | Depositories | (0.009)*** lag(1) | (0.695) | (0.015)** lag(1) | (0.105) |
| | Broker-Dealers | (0.001)*** lag(1) | (0.305) | (0.001)*** lag(1) | (0.588) |
| | Insurance Com. | (0.005)*** lag(1) | (0.502) | (0.046)** lag(2) | (0.015)** lag(2) |
| | Others | (0.000)*** lag(1) | (0.218) | (0.094)* lag(1) | (0.016)** lag(1) |
| *NoD* | Depositories | (0.001)*** lag(6) | (0.000)*** lag(6) | (0.912) | (0.005)*** lag(1) |
| | Broker-Dealers | (0.009)*** lag(1) | (0.547) | (0.026)** lag(2) | (0.986) |
| | Insurance Com. | (0.000)*** lag(5) | (0.188) | (0.014)** lag(2) | (0.039)** lag(2) |
| | Others | (0.000)*** lag(1) | (0.111) | (0.146) | (0.943) |
| *PIR* | Depositories | (0.047)** lag(2) | (0.802) | (0.249) | (0.001)*** lag(2) |
| | Broker-Dealers | (0.000)*** lag(1) | (0.216) | (0.366) | (0.325) |
| | Insurance Com. | (0.000)*** lag(2) | (0.063)* lag(2) | (0.017)** lag(2) | (0.104) |
| | Others | (0.000)*** lag(1) | (0.249) | (0.718) | (0.167) |
| Panel B: first differences | | | | | |
| *DD* | Depositories | (0.014)** lag(1) | (0.342) | (0.450) | (0.017)** lag(1) |
| | Broker-Dealers | (0.013)** lag(2) | (0.119) | (0.006)*** lag(1) | (0.909) |
| | Insurance Com. | (0.529) | (0.590) | (0.024)** lag(1) | (0.211) |
| | Others | (0.002)*** lag(2) | (0.199) | (0.798) | (0.177) |
| *NoD* | Depositories | (0.045)** lag(2) | (0.000)*** lag(2) | (0.501) | (0.443) |
| | Broker-Dealers | (0.034)** lag(1) | (0.696) | (0.009)*** lag(1) | (0.943) |
| | Insurance Com. | (0.000)*** lag(4) | (0.088)* lag(4) | (0.017)** lag(4) | (0.117) |
| | Others | (0.050)** lag(1) | (0.852) | (0.880) | (0.922) |
| *PIR* | Depositories | (0.025)** lag(1) | (0.776) | (0.378) | (0.025)** lag(1) |
| | Broker-Dealers | (0.000)*** lag(2) | (0.289) | (0.239) | (0.344) |
| | Insurance Com. | (0.109) | (0.115) | (0.009)*** lag(1) | (0.511) |
| | Others | (0.021)** lag(3) | (0.405) | (0.736) | (0.441) |

**Table 5. Predictive power**

To compare predictive ability for STLFSI, we first run a regression, in which the explanatory variables are the lagged terms of STLFSI and of the benchmark, as shown in Equation (25). Next we include the lagged terms of the full model to check for its incremental predictive ability, as shown in Equation (26). With an *F-test* (Equation (27)), we examine if the difference in $R^2$ values across the two regressions differs significantly from 0. Bold font reveals cases where $R^2$ is higher in Equation (26) than in Equation (25).

***, **, and * are significant at 1%, 5%, and 10% levels, respectively.

| | | $R^2$ Eq. (25) | $R^2$ Eq. (26) | *F-statistics* | *P-value* |
|---|---|---|---|---|---|
| Panel A: levels | | | | | |
| *DD* | Depositories | 0.917 | **0.921** | 7.761 *** | 0.006 |
| | Broker-Dealers | 0.918 | **0.919** | 0.855 | 0.358 |
| | **Insurance Companies** | 0.917 | **0.921** | 4.800 *** | 0.009 |
| | Others | 0.918 | 0.918 | 0.991 | 0.321 |
| *NoD* | Depositories | 0.922 | **0.923** | 1.963 | 0.163 |
| | Broker-Dealers | 0.918 | 0.918 | 0.401 | 0.527 |
| | Insurance Companies | 0.938 | 0.938 | 0.998 | 0.319 |
| | Others | 0.917 | **0.920** | 6.755 ** | 0.010 |
| *PIR* | Depositories | 0.917 | 0.917 | 0.000 | 1.000 |
| | **Broker-Dealers** | 0.918 | **0.920** | 5.225 ** | 0.024 |
| | **Insurance Companies** | 0.919 | **0.923** | 9.125 *** | 0.003 |
| | Others | 0.917 | **0.918** | 2.227 | 0.137 |
| Panel B: first differences | | | | | |
| *DD* | Depositories | 0.060 | **0.063** | 0.530 | 0.467 |
| | Broker-Dealers | 0.071 | **0.078** | 1.344 | 0.248 |
| | **Insurance Companies** | 0.068 | **0.102** | 6.494 ** | 0.012 |
| | Others | 0.063 | 0.063 | 0.007 | 0.931 |
| *NoD* | Depositories | 0.083 | **0.110** | 5.281 ** | 0.023 |
| | Broker-Dealers | 0.070 | **0.092** | 4.176 ** | 0.043 |
| | Insurance Companies | 0.307 | **0.312** | 1.303 | 0.255 |
| | Others | 0.062 | 0.062 | 0.010 | 0.922 |
| *PIR* | Depositories | 0.069 | **0.072** | 0.680 | 0.411 |
| | **Broker-Dealers** | 0.060 | **0.113** | 10.453 *** | 0.002 |
| | **Insurance Companies** | 0.063 | **0.336** | 71.949 *** | 0.000 |
| | Others | 0.061 | **0.065** | 0.705 | 0.402 |

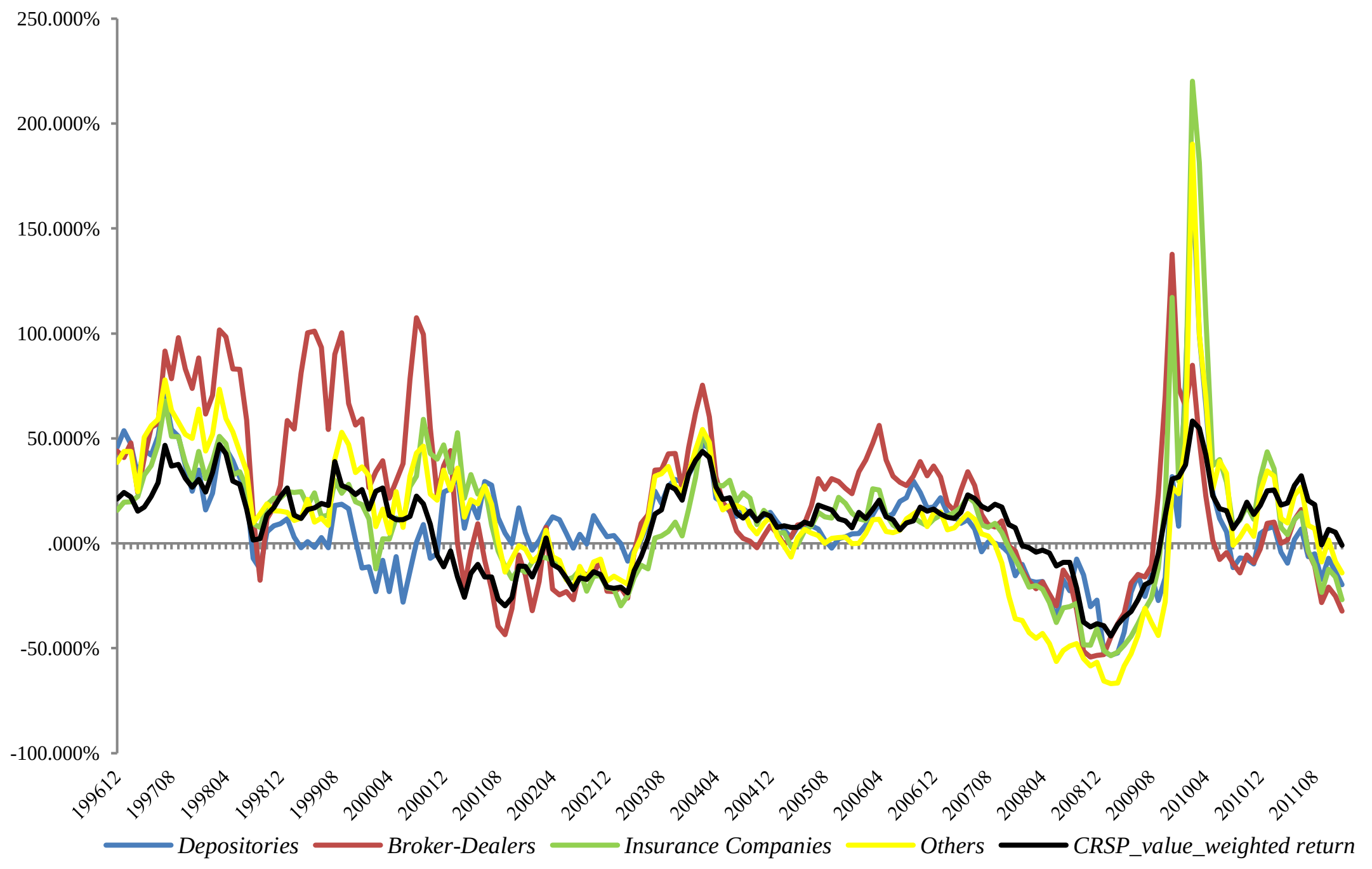

**Figure 1.** Annual equity returns by sector.

The annual equity returns, calculated by summing daily returns over the past year at the end of every month, spanned December 1996 to December 2011 for each sector. The sector-level annual return ending at month $t$ for sector $k$ is calculated by $r_{k,t} = \sum_{j=1}^{10} w_{j,k,t} \times r_{j,k,t}$, where $r_{j,k,t}$ is firm $j$'s annual return, and $w_{j,k,t}$ is the weight, based on the market equity of firm $j$ at the end of month $t$. The blue, red, green, and yellow lines represent Depositories, Broker-Dealers, Insurance Companies, and Others respectively. The black line represents the annual return on CRSP value-weighted index.

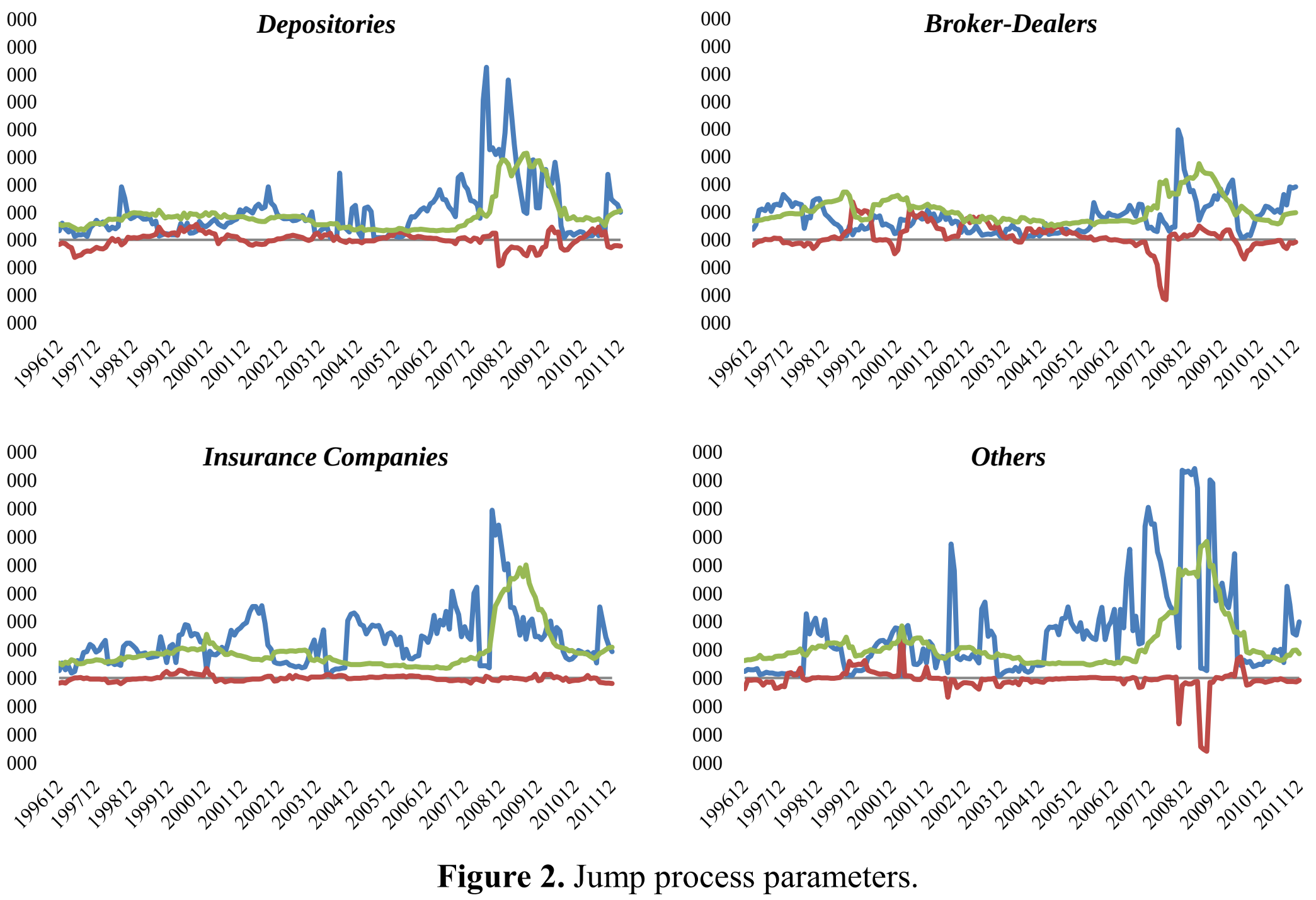

**Figure 2.** Jump process parameters.

The figure shows the three time series obtained from the estimation of correlated jumps in each sector. The results are plotted at the end of each rolling window sample, with 181 monthly observations for each time series. The blue, red, and green lines represent $\lambda$ (intensity of correlated jumps), *mu_coj* (average value of the means of jump size for 10 big financial institutions), and *std_coj* (average value of standard deviations of jump size for 10 big financial institutions), respectively.

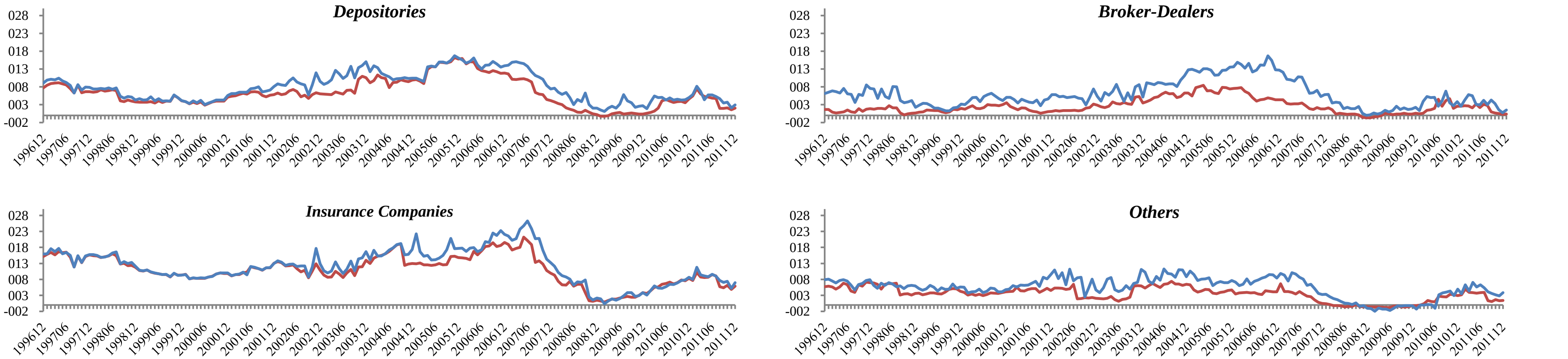

**Panel A: *DD***

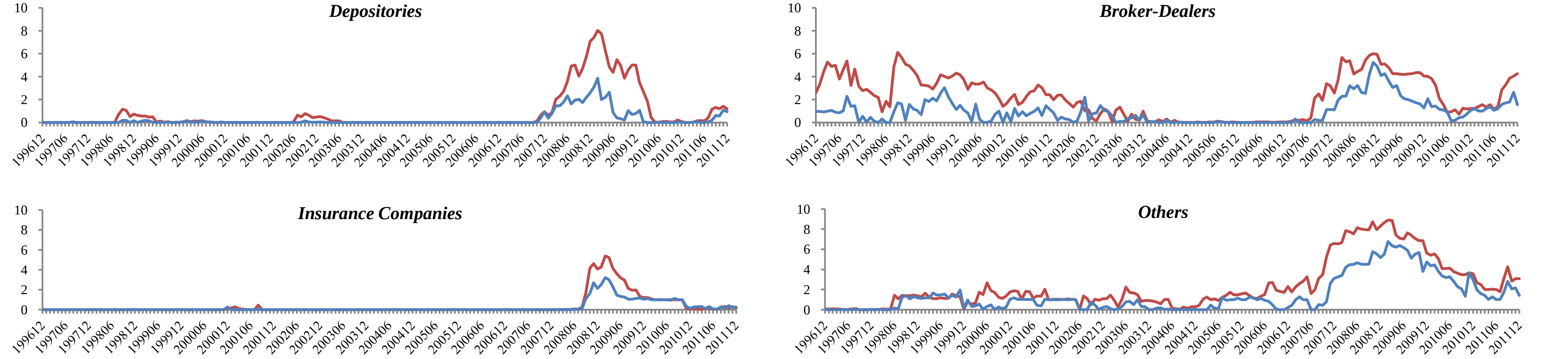

**Panel B: *NoD***

**Figure 3**. Systemic risk measures.

This figure plots three alternative systemic risk measures, *DD* (distance-to-default, Panel A), *NoD* (number of joint defaults, Panel B), and *PIR* (price of insurance ratio, Panel C) during 1996–2011 by sector. The red and blue lines represent measures derived from our model and the benchmark, respectively.. (Conti.

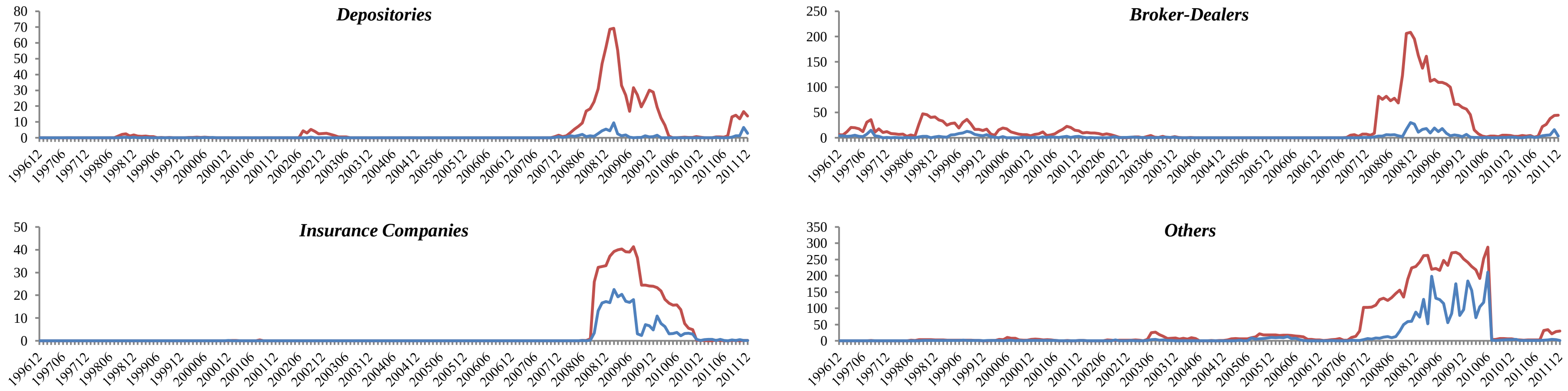

**Panel C: *PIR***

**Figure 3 (Conti.)**. Systemic risk measures.

This figure plots an alternative systemic risk measure, *PIR*, during 1996-2011, by sector. The *PIR* (Panel C) is the ratio of the price of insurance against financial distress to the aggregate asset value (scaled by multiplying $10^6$).

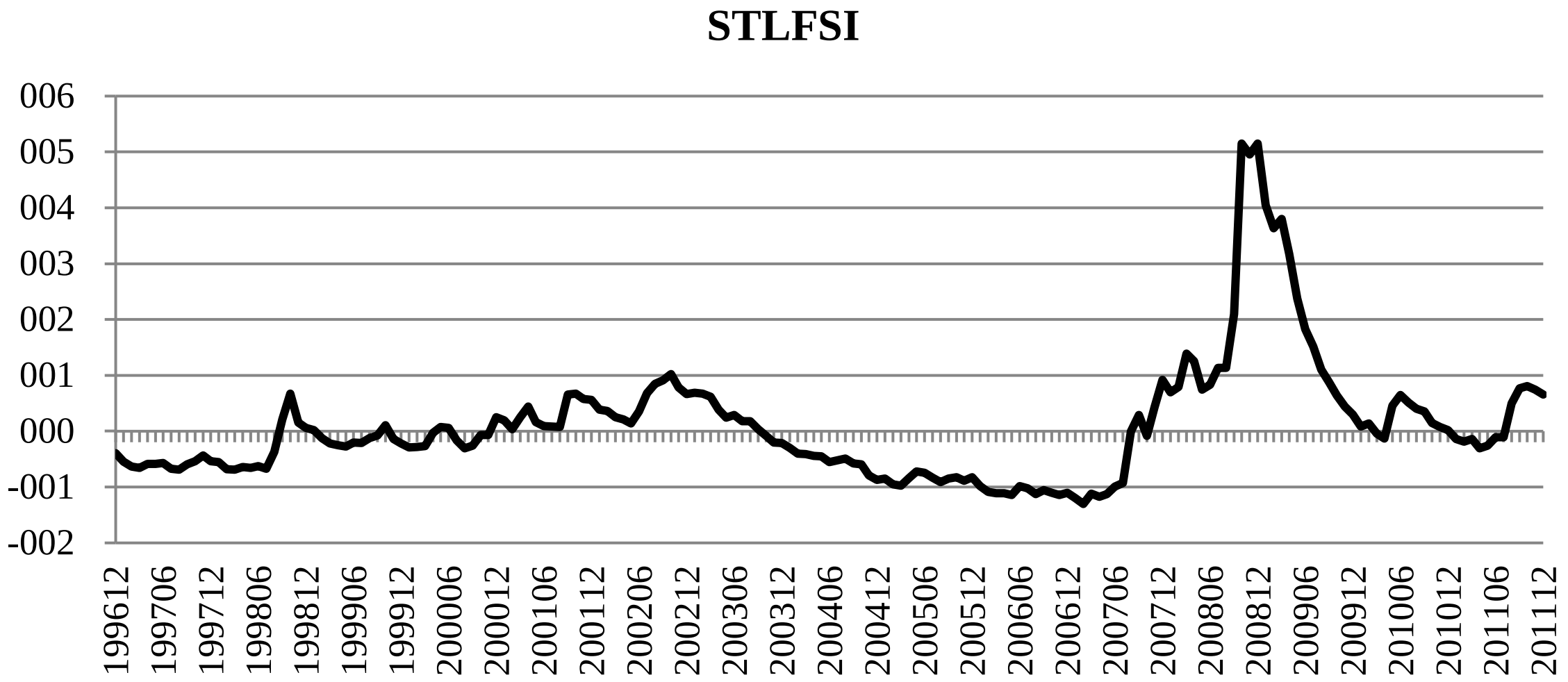

**Figure 4**. St. Louis Fed Financial Stress Index (STLFSI).
This figure contains monthly data of STLFSI, from December 1996 to December 2011.